\documentclass[%
 aps,
 amsmath,amssymb,
 reprint,
]{revtex4-1}
\usepackage[utf8]{inputenc}
\usepackage{graphicx}
\usepackage{enumitem}
\usepackage{bm}
\usepackage{cases}
\usepackage{upgreek}
\usepackage[usenames, dvipsnames]{color}
\usepackage{hyperref}
\usepackage{braket}
\usepackage[normalem]{ulem}

\usepackage{textgreek}
\usepackage{gensymb}
\usepackage{dcolumn}

\newcommand \mc[1] { \mathcal{#1} }
\newcommand \dd[1]  { \,\textrm d{#1} }   
\newcommand \rmm[1]  { \textrm{#1} }

\makeatletter
\def\@email#1#2{%
 \endgroup
 \patchcmd{\titleblock@produce}
  {\frontmatter@RRAPformat}
  {\frontmatter@RRAPformat{\produce@RRAP{*#1\href{mailto:#2}{#2}}}\frontmatter@RRAPformat}
  {}{}
}%
\makeatother
\begin{document}

\preprint{AIP/123-QED}

\title{Compact and complete description of non-Markovian dynamics}

\author{Thomas Sayer}
\affiliation{Department of Chemistry, University of Colorado Boulder, Boulder, CO 80309, USA}

\author{Andr\'{e}s Montoya-Castillo}
\homepage{Andres.MontoyaCastillo@colorado.edu}
\affiliation{Department of Chemistry, University of Colorado Boulder, Boulder, CO 80309, USA} 


\date{\today}

\begin{abstract}

Generalized master equations provide a theoretically rigorous framework to capture the dynamics of processes ranging from energy harvesting in plants and photovoltaic devices, to qubit decoherence in quantum technologies, and even protein folding. At their center is the concept of memory. The explicit time-nonlocal description of memory is both protracted and elaborate. When physical intuition is at a premium one would desire a more compact, yet complete, description. Here, we demonstrate how and when the time-convolutionless formalism constitutes such a description. In particular, by focusing on the dissipative dynamics of the spin-boson and Frenkel exciton models, we show how to: easily construct the time-local generator from reference reduced dynamics, elucidate the dependence of its existence on the system parameters and the choice of reduced observables, identify the physical origin of its apparent divergences, and offer analysis tools to diagnose their severity and circumvent their deleterious effects. We demonstrate that, when applicable, the time-local approach requires as little information as the more commonly used time-nonlocal scheme, with the important advantages of providing a more compact description, greater algorithmic simplicity, and physical interpretability. We conclude by introducing the discrete-time analogue  and a straightforward protocol to employ it in cases where the reference dynamics have limited resolution. The insights we present here offer the potential for extending the reach of dynamical methods, reducing both their cost and conceptual complexity.

\end{abstract}

\maketitle

\section{Introduction}

Predicting the dynamics of large, complex many-body systems stands as a grand theoretical challenge which holds the key to addressing problems ranging from protein folding and glass formation, to energy harvesting in plants and artificial devices and electronic and thermal transport, and quantum information protection and processing. In quantum systems the exponential growth in the cost of the simulations with both system size and simulation time exacerbate the difficulty of predicting dynamics over long timescales. Thus, a general framework to capture the equilibrium and non-equilibrium dynamics of such complex systems is urgently needed. 

Generalized master equations (GMEs) \cite{BreuerPetruccione} offer the means to overcome many of these complexities. Their advantage arises from the realization that one is often not interested in the dynamics of the entire system but rather the dynamics of a small set of observables, such as the transition rates among a handful of protein configurations, the evolution of a few electronic excitations in molecular aggregates, or the non-equilibrium state of a qubit subject to interactions with a complex environment. In such cases, projection operator techniques \cite{Grabert1982, FickSauermann, BookZwanzig} allow one to rigorously derive a low-dimensional equation of motion for the observables of interest. This reduction in dimensionality, however, comes at a cost: the reduced dynamics become non-Markovian and the correlation between the desired observables and the rest of the system is encoded into a memory term, whose solution is as difficult as solving the original problem. For simplicity of language, we henceforth refer to the desired observables as the projected observables and the rest of the system as the orthogonal subspace. Given their generality and versatility, GMEs have proven critical to predict, elucidate, and analyze a wide variety of phenomena, including human breast cancer migration \cite{Mitterwallner2020}, the slow configurational dynamics of protein folding \cite{CaoMontoya2020, Ayaz2021, Zhu2021d, Unarta2021a, Ayaz2021a}, the development of markers of neurological diseases \cite{Demin2008}, structural relaxation in polymers \cite{Felderhof2008, Schweizer1998}, non-equilibrium exciton and charge transfer in molecular systems and driven transport in nanoscopic devices \cite{MayKuhn, NitzanBook, Wilner2013, Kidon2015a, Granger2012, Schinabeck2020}, qubit decoherence \cite{Shabani2005, Vacchini2010, Barnes2012, Ng2022}, density fluctuations in glasses \cite{BookGoetze, Reichman2001, Janssen2015}, classical, quantum, and relativistic hydrodynamics \cite{BookBoonYip, BookHansenMcdonald, Koide2009, Huang2011, Hartnoll2012, Lucas2016, Han2021}, the general relativistic dynamics of the cosmic jerk \cite{TeVrugt2021}, and even fluctuations in market behavior \cite{Picozzi2002, Meng2016}.

GMEs come in time-convolutional (TC) \cite{Nakajima1958, Mori1958, Zwanzig1960a} and time-convolutionless (TCL) \cite{Tokiyama1976, Chaturvedi1979} forms. In the TC approach, the memory kernel, $\mc{K}(t)$, encodes the non-Markovian evolution of the projected observables and, through its convolution over the entire history of the dynamics, produces the projected equation of motion (see Eq.~(\ref{eq:TCGME})). Crucially, when the projected observables are the slowest degrees of freedom in the entire system and the dynamics are sufficiently dissipative, $\mc{K}(t)$ decays on timescales shorter than those associated with the projected dynamics. In such cases, calculating the memory kernels instead of the full projected dynamics can greatly improve computational efficiency. Recently, this has inspired many successful approaches \cite{Shi2003, Zhang2006a, Cohen2013, Ivanov2015, KellyMontoya2016, Montoya2016a, Montoya2017b, Mulvihill2021b, Mulvihill2022, Ng2021}. Nevertheless, while this approach provides a complete computational framework to predict non-Markovian dynamics, extracting physical insight from these memory kernels is challenging and often requires expert knowledge. 

The TCL formulation, on the other hand, encodes the non-Markovian evolution of the projected observables in a time-dependent rate matrix or time-local generator, $\mc{R}(t)$. Like $\mc{K}(t)$, $\mc{R}(t)$ can have a short equilibration timescale offering similar efficiency benefits as the TC-GME while offering a more transparent and easier to evolve form of the reduced equation of motion (see Eq.~(\ref{eq:TCL-GME})). While recent advances have shown how to successfully calculate the time-local generator for specific systems \cite{Nan2009a, Kidon2015a, Kidon2018}, other works show that the time-local generator can suffer from divergences that prevent its use \cite{LiuShi2018, Kropf2016, Maldonado-Mundo2012}. Hence, fundamental questions remain: 
\begin{enumerate}
    \item Is there a simple way to construct the time-local generator that is agnostic to the system and the underlying dynamics (classical, quantum, or relativistic)?
    \item Since the TCL formulation has strict existence requirements such that a time-local description may not always be possible, can one develop insights into when the TCL form exists and elucidate the factors on which its existence depends?
    \item When the TCL form exists, does its time-local generator, $\mc{R}(t)$, decay on a timescale similar to that of the memory kernel $\mc{K}(t)$ in the TC-GME? 
    \item Is it possible to establish a discrete-time analog of the TCL-GME that one can use even in cases where the reference dynamics have poor temporal resolution or are beset by noise?
\end{enumerate}

Here, we address the above questions. In particular, we develop a simple, accurate, and efficient means to exploit the advantages of TCL-GMEs to provide a highly compact and complete representation of the non-Markovian dynamics governing projected observables and thereby reduce the computational cost and extend the applicability of both classical and quantum dynamics \cite{Tanimura1989, Suess2014, Varvelo2021, Makri1995a, Makri2020, Prior2010, Tamascelli2019, Strathearn2018, Cygorek2022, Werner2006, Gull2011, Cohen2015, Meyer1990, Wang2003, White2004, Vidal2004, Daley2004} approaches. We show that the TCL approach requires as little reference dynamics to construct as the TC scheme---which has already been demonstrated to provide a compact means to encode non-Markovian dynamics of various systems \cite{Shi2003, Cohen2011, Cohen2013, Cohen2013a, Wilner2013, Wilner2014, Pfalzgraff2015, KellyMontoya2016, Montoya2016a, Montoya2017b, Pfalzgraff2019, Mulvihill2021b, Mulvihill2022, Ng2021, Cerrillo2014, Pollock2018a, Joergensen2020, Rosenbach2016, Kananenka2016, Pollock2018, Carof2014, Lesnicki2016, CaoMontoya2020}---while entirely avoiding the complexities of the time-nonlocal convolution over the memory kernel and its construction. Admittedly, as suggested above, it has long been appreciated \cite{BreuerPetruccione} that a time-local description is not guaranteed to exist for all problems due to the appearance of mathematical divergences in the time-local generator. However, previous works have largely focused on a perturbative treatment of the time-local generator, for which the appearance of poles precludes further investigation \cite{LiuShi2018, Nestmann2021}. It would stand to reason that this pathology remains unworkable in the exact description. Yet, anticipating our results, we show that not only can many poles be removed by a change of projector, most are inconsequential and can be outright ignored. We further demonstrate that even in regimes where the exact time-local generator does not in fact exist, an approximate description remains remarkably well-behaved. Below, we employ the spin-boson (SB) model \cite{Leggett1987, Weiss} and the Frenkel exciton model \cite{MayKuhn} parameterized for the Fenna-Matthews-Olson (FMO) complex \cite{Adolphs2006, Ishizaki2009b} to develop, analyze, and illustrate our approach. While the TCL-GME is a continuous-time scheme, we also provide the straightforward steps needed to generalize the TCL-GME to be able to treat discrete-time data, such as obtained from methods where high temporal resolution is expensive or impossible. Importantly, our discrete-time TCL-GME offers a highly efficient way to capture the non-Markovian propagator and determine the onset of Markovianity while circumventing the need for time-derivatives of the reference data. We emphasize that these methods apply equally to classical problems such as protein folding, where the extension to low-resolution data is of particular interest. 

\section{Formal dimensionality reduction} 

Whether from an experimental or theoretical perspective, one is often interested in a small subset of observables, such as nonequilibrium averages or equilibrium time correlation functions. These dynamical objects underlie our description of spectroscopy, transport, and chemical reaction kinetics. The projection operator formalism \cite{Grabert1982, FickSauermann, BookZwanzig} provides a rigorous and convenient means to derive an equation of motion for this set of observables---the GME. At the heart of this formalism are the projection operators: $\mc{P} \equiv |\mathbf{A})(\mathbf{A}| = \sum_{j=1}^{N}|A_j)(A_j|$, which defines the $N$ observables whose dynamics one is pursuing, and $\mc{Q} \equiv 1 - \mc{P}$, which encompasses the ``uninteresting'' orthogonal subspace. The inner product used to define the projector, $(A|B)$, should be constructed to best suit the needs of the problem. For example, when one is interested in equilibrium-time correlation functions to obtain transport coefficients \cite{KuboStatMechII}, chemical reaction rates \cite{Yamamoto1960, Miller1983a}, or spectroscopic responses \cite{Mukamel-Book}, one can use a Mori-type projector in the form of the Kubo-transformed correlation function or the symmetrized correlation function \cite{KuboStatMechII, FickSauermann, Montoya2017b}. Alternatively, one can choose a non-equilibrium projector \cite{Argyres1964, Sparpaglione1987} that tracks the dynamics of all or a subset of states.

By encoding the exact dynamical correlation between the projected observables and the orthogonal subspace into a low-dimensional equation of motion, GMEs provide an economical way of describing the non-Markovian dynamics of the projected observables. To date, most work on the exact calculation of non-Markovian dynamics in reduced subspaces employs the TC-GME \cite{Nakajima1958, Mori1958, Zwanzig1960a},
\begin{equation}\label{eq:TCGME}
    \dot{\mc{C}}(t) = \dot{\mc{C}}(0) \mc{C}(t) - \int_0^{t}d\tau\ \mc{K}(\tau)\mc{C}(t-\tau),
\end{equation}
which encodes the dynamical correlation between the projected observables and orthogonal subspace in a time-nonlocal memory term $\mc{K}(t)$. Finding $\mc{K}(t)$ is crucial in employing the TC-GME. \vfill\pagebreak
Here
\begin{equation}
    \mc{C}(t) = (\mathbf{A}| e^{-\alpha \mathcal{L}t} |\mathbf{A})
\end{equation}
corresponds to the non-equilibrium average or equilibrium time correlation function of the projected observables. We note that the GME framework is general and applies equally well to both classical and quantum systems. Specifically, for the quantum case, $\alpha = \pm i$ and $\mathcal{L} \equiv [\hat{H}, \cdot]$ is the quantum commutator, whereas in the classical case, $\alpha = \pm 1$ and $\mathcal{L} \equiv \{H, \cdot \}_{\rm PB}$ is the classical Poisson bracket. The $\pm$ in the definition of $\alpha$ allows one to adopt the Schrodinger or Heisenberg pictures. A discrete-time analogue of the TC-GME, the Tensor Transfer Method~(TTM) \cite{Cerrillo2014, Pollock2018a, Joergensen2020}, can be derived and has served as a complementary technique in capturing the dynamics of projected systems \cite{Rosenbach2016, Kananenka2016, Pollock2018}.

Yet, questions remain as to whether the TC-GME provides the most straightforward and efficient approach to non-Markovian dynamics. While recent work has provided expressions $\mc{K}(t)$ in terms of additional measurements of correlation functions \cite{Shi2003, Zhang2006a, Cohen2013, Ivanov2015, KellyMontoya2016, Montoya2016a, Montoya2017b, Mulvihill2021b, Mulvihill2022, Ng2021}, these, at best, require the first and second derivatives of $\mc{C}(t)$, the solution of a Volterra equation of the second kind to extract the memory kernel, and subsequently the solution of the integro-differential equation that is the TC-GME. Even the discrete-time TTM requires nontrivial unfolding of the reduced dynamics to construct the memory tensor.

We seek a highly compact and complete description of the non-Markovian dynamics of reduced observables, motivating the use of equations that only require a small amount of reference data to construct a non-Markovian generator and are also local in time. Hence, we eschew time-nonlocal TC-GMEs and TTMs and instead turn to the time-local TCL-GME \cite{Tokiyama1976, Chaturvedi1979},
\begin{equation}\label{eq:TCL-GME}
    \dot{\mc{C}}(t) = \mc{R}(t)\mc{C}(t),
\end{equation}
where the time-local generator takes the form
\begin{equation}\label{eq:R-definition}
    \mc{R}(t) = -\alpha (\mathbf{A}|\mc{L}[1-\Sigma(t)]^{-1}|\mathbf{A}),
\end{equation}
and $\Sigma(t) = \int_{0}^t\dd{s}\,e^{-\alpha\mc{Q}\mc{L}s}\mc{Q}(-\alpha \mc{L})\mc{P}e^{-\alpha\mc{L}s}$.  
Clearly, equation~(\ref{eq:R-definition}) is only valid when the inverse of $1 - \Sigma(t)$ is well defined. Additionally, finding $\mc{R}(t)$ in the TCL-GME, just like finding $\mc{K}(t)$ in the TC-GME, is as difficult as directly solving the original quantum dynamics problem. Despite this difficulty, both GMEs have served as starting points for approximate treatments, including perturbative expansions (including in the Markovian limit) \cite{Bloch1957, Redfield1965, Leggett1987, Dekker1987, Zhang1998a, Golosov2001d, Cheng2005, Jang2008, Kolli2011}, assumed functional forms \cite{Harp1970, Chen2010a}, and hierarchical expansions truncated by assuming a closure based on, say, Gaussian statistics \cite{Tanimura1989, Ishizaki2009, Zaccarelli2001, Janssen2015}.  

More recently, the TCL-GME has been obtained with high accuracy using exact methods for various systems \cite{Nan2009a, Kidon2015a, Kidon2018, LiuShi2018}. Here, we obtain $\mc{R}(t)$ directly by left-multiplying Eq.~(\ref{eq:TCL-GME}) on both sides by the inverse of $\mc{C}(t)$, yielding,
\begin{equation}\label{eq:R-from-inverse}
    \mc{R}(t) = \dot{\mc{C}}(t)[\mc{C}(t)]^{-1}.
\end{equation}
Unlike previous approaches, we do not rely on the microscopic definition of $\mc{R}(t)$ in Eq.~(\ref{eq:R-definition}). Instead, we assert that a time-local description of the non-Markovian dynamics of projected variables exists and that the time-local generator $\mc{R}(t)$ is given by Eq.~(\ref{eq:R-from-inverse}), which requires only the invertibility of the projected observables, $\mc{C}(t)$. We note that our approach to obtain $\mc{R}(t)$ in Eq.~(\ref{eq:R-from-inverse}) is analogous to that introduced in Refs.~\onlinecite{Kidon2015a, Kidon2018, LiuShi2018}. The main point of contrast is that we do not require an explicit expression for the time-local non-Markovian propagator and its time derivative, but rather construct $\mc{R}(t)$ directly from the correlation matrix $\mc{C}(t)$ and its time derivative instead. Indeed we later show that one can avoid a time-derivative altogether by employing an entirely discrete-time approach (see Sec.~\ref{sec:discrete-time-tcgme}). Obtaining the time-local generator in this manner directly reflects the fact that $\mc{R}(t)$ governs all initial conditions that span the projection operator, rather than just one initial condition corresponding to a single column (or a normalized linear combination of the columns) of $\mc{C}(t)$.

It initially seems from Eq.~(\ref{eq:R-from-inverse}) that $\mc{R}(t)$ is as long-lived as $\mc{C}(t)$. A similar criticism could be levelled at the TC-GME, Eq.~(\ref{eq:TCGME}), (and the discrete-time TTM) where the projected observables are convolved with $\mc{K}(t)$. Yet, when the projected observables evolve on a slower timescale than the orthogonal space the memory kernel is found to decay to zero over a short timescale, $\tau_K$, i.e., $\mc{K}(t > \tau_K) = 0$. A short-lived $\mc{K}(t)$ suggests that one need only use the short-time dynamics of $\mc{C}(t)$ to fully determine its entire evolution, leading to significant gains in efficiency. In fact, this short lifetime for $\mc{K}(t)$ underlies the success of TC-GMEs in ameliorating and delaying the onset of the dynamical sign problem in real-time path integral Monte Carlo \cite{Cohen2011}, capturing the long-time exciton dynamics in large light harvesting complexes \cite{Pfalzgraff2019, Mulvihill2021}, and interrogating the uniqueness of steady states in driven nonequilibrium junctions \cite{Wilner2013, Wilner2014}. For the TCL-GME, this separation of timescales leads the non-Markovian generator, $\mc{R}(t)$, to reach its long-time limit over a short timescale, $\tau_R$, beyond which $\mc{R}(t > \tau_R)$ is a constant matrix. In such cases, only short-time reference data is necessary to capture $\mc{R}(t)$, offering the potential for significant cost reduction in the simulation of $\mc{C}(t)$ which extends over arbitrary timescales. 

\section{Continuous-time TCL-GME}

In the following section, we use the spin-boson (SB) and many-level Frenkel exciton models to interrogate the ability of the TCL-GME to efficiently and accurately capture the dynamics of projected observables. In particular, we assess the feasibility of using Eq.~(\ref{eq:R-from-inverse}) to construct the time-local generator, $\mc{R}(t)$. Importantly, we examine when $\mc{R}(t)$ exists, elucidate its dependence on the Hamiltonian parameters and projection operator, and analyze the origin of spikes in $\mc{R}(t)$ and their impact on the resulting dynamics. 
\vfill\pagebreak

In all numerical demonstrations, we focus on the nonequilibrium dynamics of either the entire electronic reduced density matrix or a subset of its elements, subject to a nonequilibrium initial condition where the electronic subsystem and bath are initially uncorrelated, 
\begin{equation}\label{eq:Liouville-dynamical-matrix}
    \mathcal{C}_{j,k}(t) = \mathrm{Tr}[A^{\dagger}_{j} e^{-i\mathcal{L}t} A_k \rho_B].
\end{equation}
Here $\rho_B = e^{-\beta H_B}/\mathrm{Tr}_B[e^{-\beta H_B}]$ is the canonical density of the bath in thermal equilibrium with the electronic ground state. The electronic operators $\{ A_j \}$ differ depending on the choice of projection operator. Here, we employ either the Argyres-Kelley \cite{Argyres1964} or the populations-based \cite{Sparpaglione1987} projection operators (see Ref.~\onlinecite{Montoya2016a} for an extended discussion of these nonequilibrium projectors). In the Argyres-Kelley projection operator, $A_j$ spans all electronic outer products, $\ket{n}\bra{m}$, giving access to the entire reduced electronic density matrix subject to all initial conditions of the product form $\rho_B \ket{n}\bra{m}$. In contrast, in the populations-based projection operator, $A_j$ spans only the electronic populations, $\ket{n}\bra{n}$, giving access only to the population dynamics subject to population-based nonequilibrium initial conditions. To obtain the reference dynamics that allows us to construct the time-local generator from Eq.~(\ref{eq:R-from-inverse}) and assess the performance of the TCL framework, we employ the numerically exact hierarchical equations of motion (HEOM) method implemented in the open-source pyrho code \cite{Berkelbach2020}. See Appendix~\ref{app:compdeets} for details.

\begin{figure}[b]
\vspace{+14pt}
\begin{center}
    \resizebox{.5\textwidth}{!}{\includegraphics{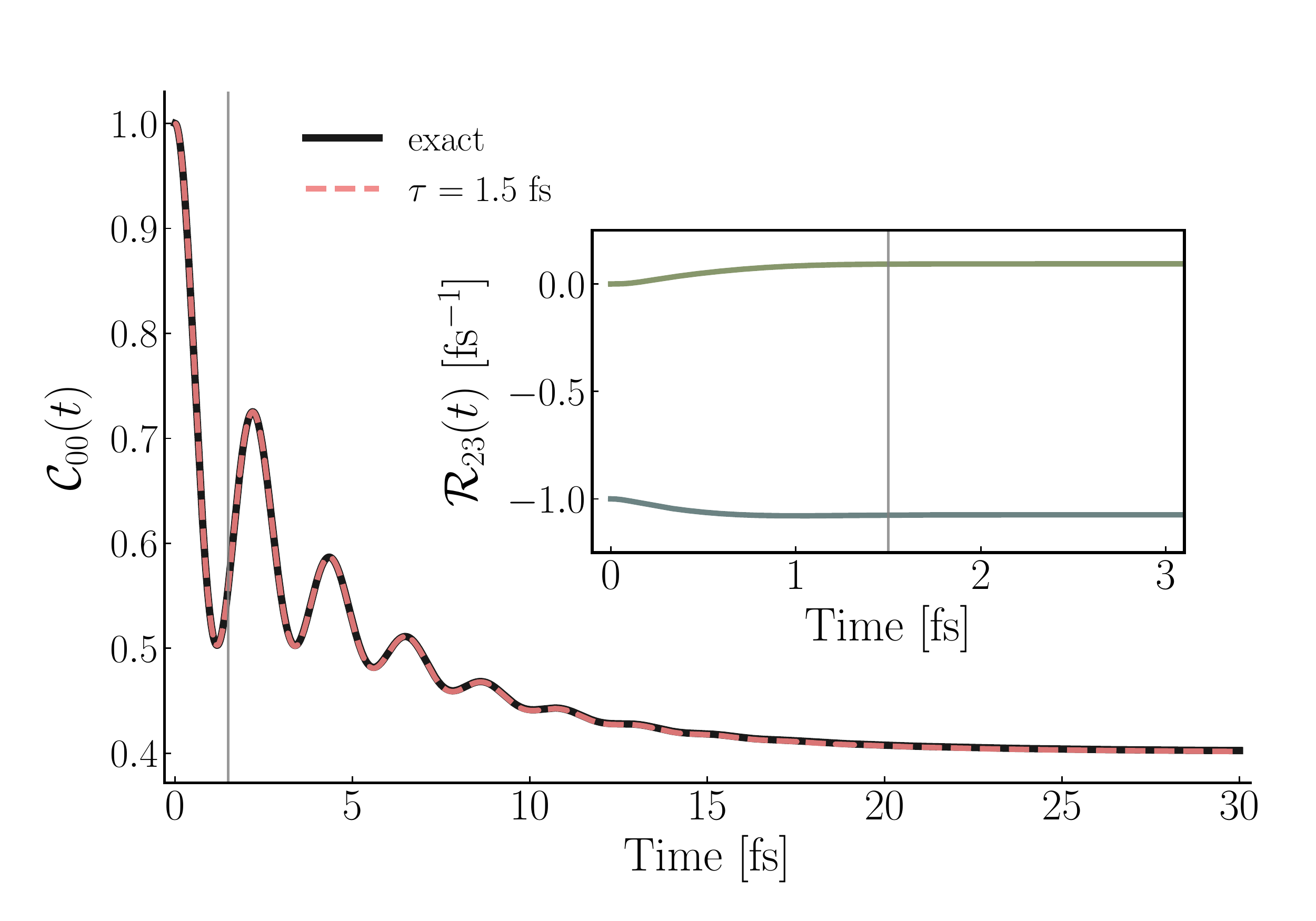}}\vspace{-2pt}
\end{center}
\vspace{-14pt}
    \caption{Population dynamics of the SB model. Parameters: $\varepsilon = 1$, $\omega_c = 5$, $\lambda = 0.1$, and $\beta = 0.2$, energy units are $\Delta$. Vertical grey lines show $\tau=1.5$~fs, when $\mc{R}(t)$ is fixed as a constant matrix. The GME dynamics from this truncated $\mc{R}(t)$ reproduce the exact dynamics to graphical accuracy. \textbf{Inset}: representative element of the time-local generator $\mc{R}(t)$, real part in green and imaginary part in teal, showing plateauing behaviour on the femtosecond timescale.}
    \label{fig:biased-SB-model-good-example}
\end{figure}

\subsection{Time-local description: existence, efficiency, and instabilities}

We begin our analysis with the SB model. In addition to being an important workhorse in simulating and elucidating fundamental features of charge and energy transfer reactions, qubit decoherence, and tunneling of light particles in metals and glasses \cite{Leggett1987, Weiss}, the SB model provides a physically transparent problem on which to benchmark our approach to the TCL-GME. Figure~(\ref{fig:biased-SB-model-good-example}) illustrates the advantages of using the TCL-GME to accurately and efficiently capture the numerically exact nonequilibrium dynamics of the SB model after an electronic excitation.  We employ the Argyres-Kelley projector \cite{Argyres1964} to focus on the spin's reduced density matrix subject to all nonequilibrium initial conditions, $\rho_k(0) = \hat{A}_k \rho_B$. For the TCL-GME to provide a more compact and complete description of the dynamics of $\mc{C}(t)$ than the TC-GME, the amount of reference dynamics required to construct $\mc{R}(t)$ needs to be comparable or smaller than that required to construct $\mc{K}(t)$. In other words, $\tau_R \leq \tau_K$. Indeed, for this system, $\tau_K = 1.5$~fs (see Appendix~\ref{app:memorykernels}, Fig.~\ref{fig:memory_kernels_SB}), confirming the view that the TCL-GME provides the most parsimonious description of these dynamics, while obviating the complications of a time-nonlocal TC-GME.

The validity of Eq.~(\ref{eq:R-from-inverse}) to obtain $\mc{R}(t)$ relies on the invertibility of $\mc{C}(t)$, which demands a description of the factors that lead to $\mc{C}(t)$ becoming non-invertible. One source of non-invertibility is that, at long times, one expects $\mc{C}(t)$ to equilibrate, leading the density matrix to approach the canonical distribution, $\rho(t \rightarrow \infty) \propto e^{- \beta H}$. This implies that different nonequilibrium initial conditions will evolve toward the same long-time limit, such that the measurement of the associated density matrix elements yield the same values at sufficiently long times, rendering two or more rows (which index the initial condition) equivalent \footnote{
One must make a distinction between traceless versus properly normalized initial conditions. While physically allowed density matrices in quantum mechanics are normalized, one can construct dynamical quantities that appear to arise from traceless initial density matrices. For example, in the Argyres-Kelley projector, which recovers the entire density matrix subject to all uncorrelated initial condition, traceless initial densities correspond to cases where the spin starts in a coherence, $\ket{j}\bra{k}$, while the bath is originally in thermal equilibrium, $\rho_B = e^{-\beta H_B}/\mathrm{Tr}_{B}[e^{-\beta H_B}]$. Physically, such a situation arises from, say, the measurement of a transition dipole operator after an impulsive initial condition. In contrast, normalized initial densities in the Argyres-Kelley projector arise from the population-based initial conditions where the spin starts from a normalized superposition of states. 
}. Since when two or more rows or columns in $\mc{C}(t)$ become equal (or proportional), as in $\mc{C}(t\rightarrow \infty)$, the matrix becomes singular, in this limit $\mc{R}(t)$ clearly cannot be obtained by inverting $\mc{C}(t)$. However, in all relevant cases, where $\mc{C}(t)$ approaches equilibrium more slowly than the orthogonal subspace, $\mc{R}(t)$ plateaus---or its structure stops mattering---before $\mc{C}(t)$ reaches equilibrium, and we avoid this non-invertible region of $\mc{C}(t\rightarrow \infty)$ altogether.

\begin{figure}[t]
\vspace{-10pt}
\begin{center}
    \resizebox{.5\textwidth}{!}{\includegraphics{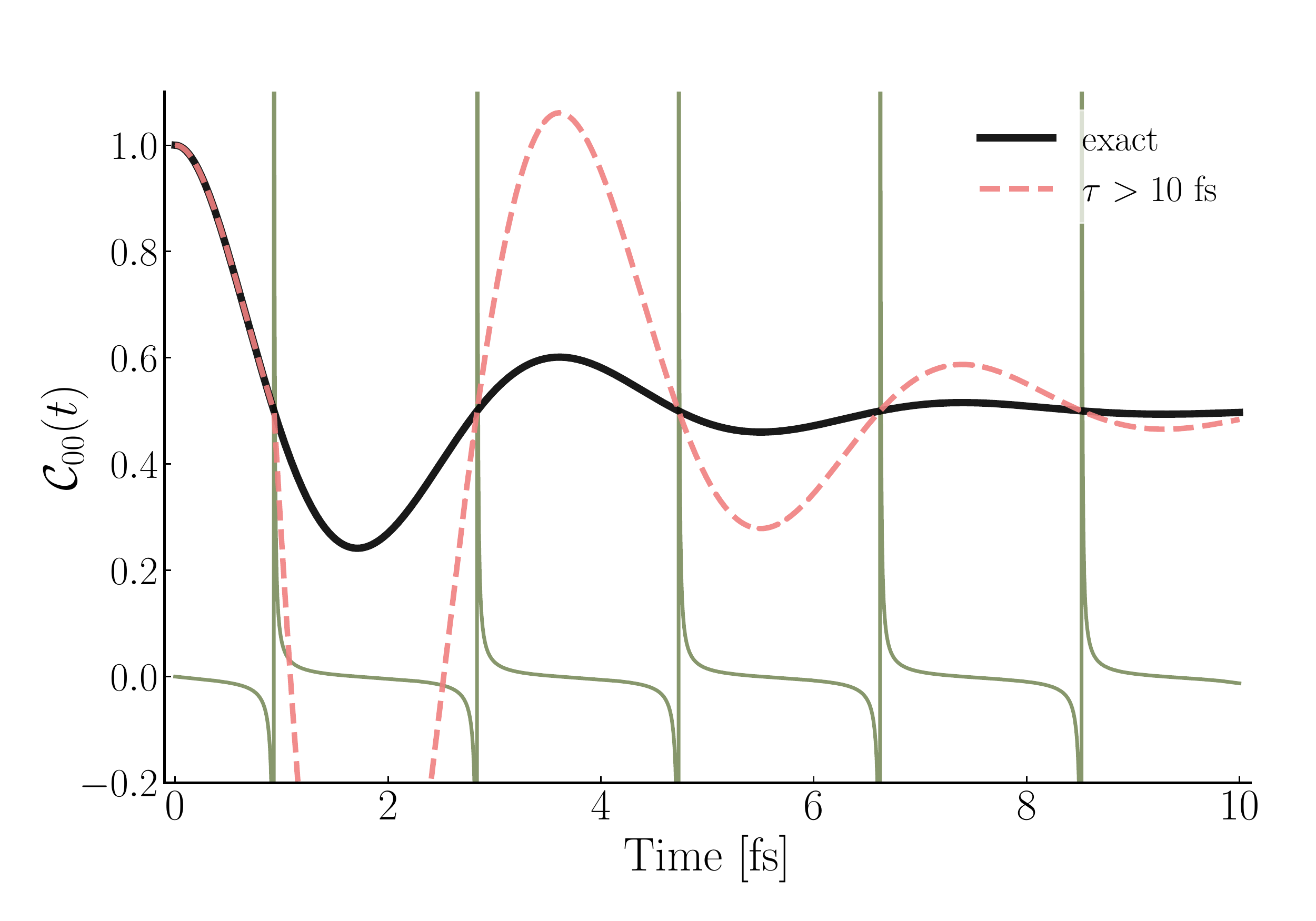}}\vspace{-2pt}
\end{center}
\vspace{-14pt}
\caption{\label{fig:popsOnly_SB} Populations-only projector with HEOM (black) and GME (dashed red) dynamics for an unbiased SB model. Parameters: $\omega_c = 2$, $\lambda = 0.1$, and $\beta = 0.1$, energy units are $\Delta$. We do not implement a cutoff for $\mc{R}(t)$ and instead use the entire interval up to $10$~fs to generate the GME dynamics. Superimposed 0.01*$\mc{R}(t)$~[fs$^{-1}$] in green exhibits spikes at regular intervals. The first divergence is sufficiently large that the numerical integration results in a discontinuous first derivative for the GME dynamics at that point (later spikes have no perceivable effect).}
\end{figure}

Before equilibrium, the appearance of spikes in $\mathcal{R}(t)$ depends on the system parameters and choice of projection operator. For instance, when using the populations-only projector \cite{Sparpaglione1987}, which captures only population dynamics with population-based initial conditions, we observe spikes in $\mc{R}(t)$ whenever the populations cross for an unbiased SB model, consistent with recent work \cite{LiuShi2018, Maldonado-Mundo2012}. As Fig.~\ref{fig:popsOnly_SB} shows, population crossings render $\mc{C}(t)$ non-invertible and lead to GME dynamics (red) that only agree with the exact dynamics (green) before the first spike in $\mc{R}(t)$ (see Appendix~\ref{app:SVD_discussion} for further discussion). This represents the worst-case scenario for the TCL-GME, indicating that for the chosen set of reduced observables a time-local description of the dynamics is formally impossible. In such cases it would appear that one must resort to the TC-GME as the most compact and complete description of the non-Markovian dynamics. Yet, as we will show in Sec.~\ref{sec:discrete-time-tcgme}, our discrete-time version of the TCL-GME allows us to resolves the issue from a numerical standpoint, demonstrating that it can still provide the most parsimonious but accurate description of the non-Markovian dynamics. 

\begin{figure}[t]
\begin{center}
    \resizebox{.5\textwidth}{!}{\includegraphics{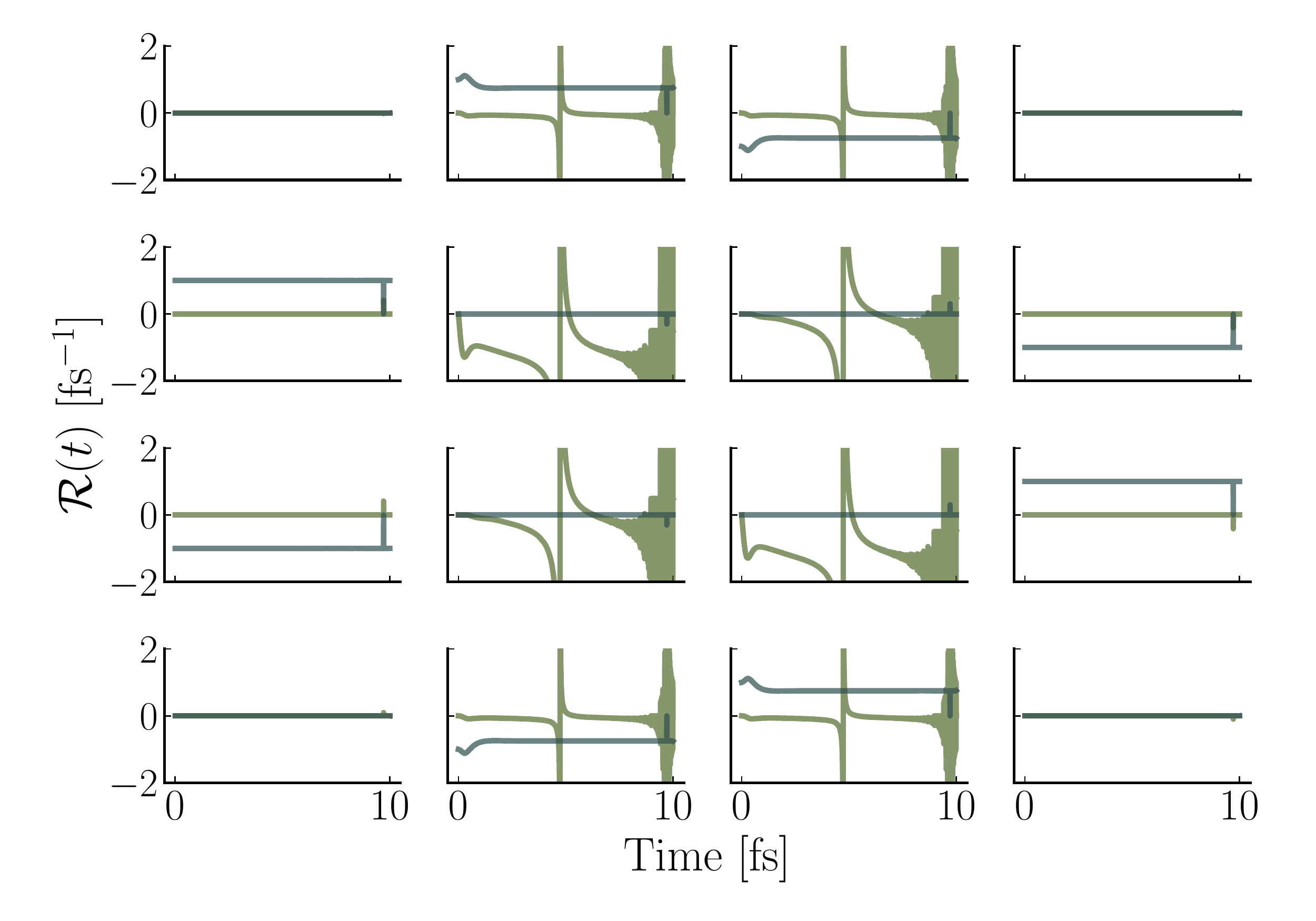}}\vspace{-2pt}
    \resizebox{.5\textwidth}{!}{\includegraphics{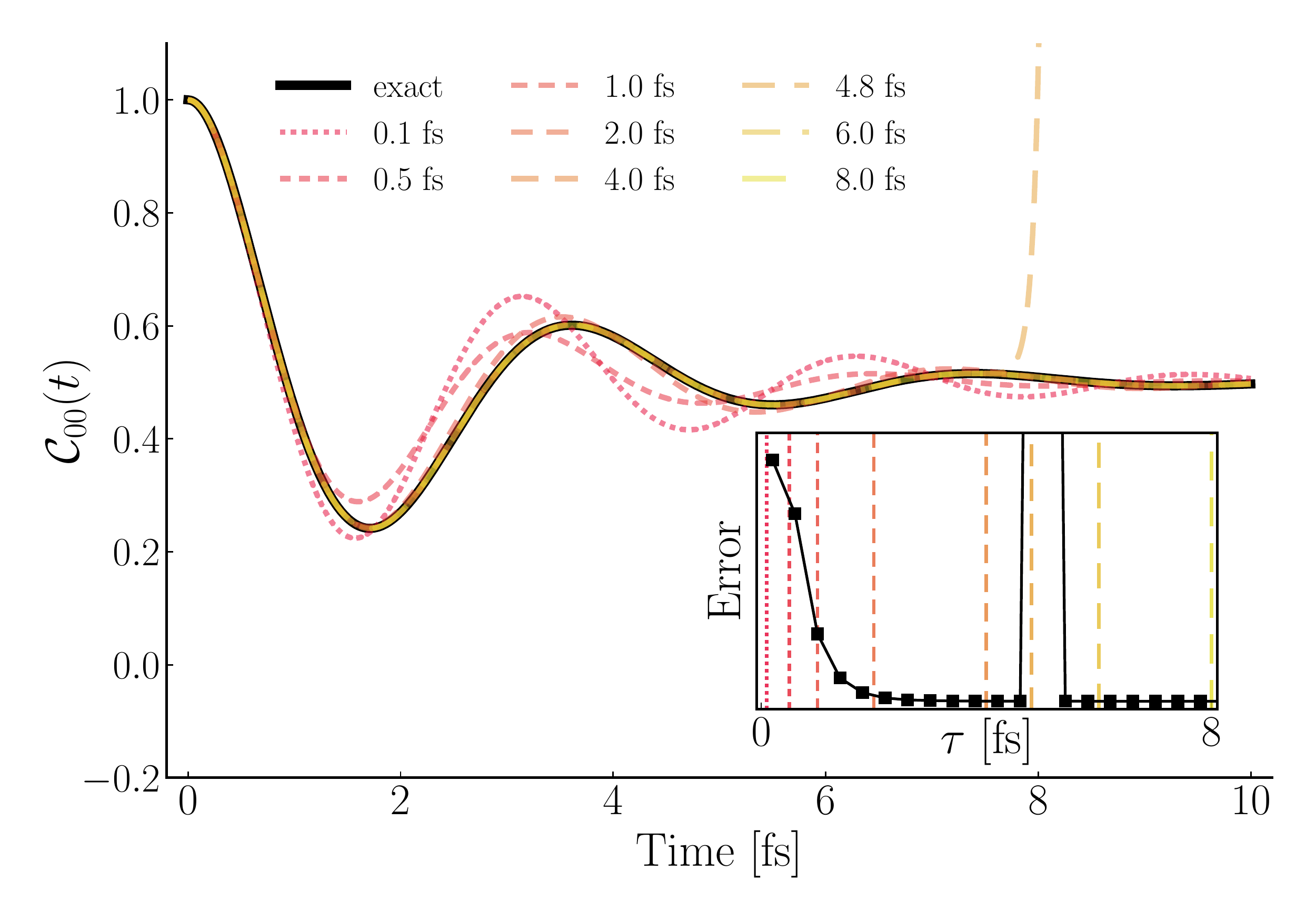}}\vspace{-2pt}
\end{center}
\vspace{-18pt}
\caption{\label{fig:CR_SB} Argyres-Kelley dynamics for an unbiased SB model. Parameters as in Fig.~\ref{fig:popsOnly_SB}. \textbf{Top}: All matrix elements of $\mc{R}(t)$ showing two columns with spikes in the real (green) part. \textbf{Bottom}: GME dynamics converging with increasing $\tau$ given in the legend, though with an instability at $\tau=4.8$~fs. \textbf{Inset}: Error between GME and exact dynamics shows $\tau_\mc{R}$ is around $4$~fs. The region of instability has a finite width before converged dynamics are once again obtained.}
\end{figure}

The observation that $\mc{R}(t)$ diverges at curve crossings when using the populations-only projector for the SB model motivates the question: do similar divergences manifest in $\mc{R}(t)$ for a different projector, such as that which yields the full reduced density matrix or other dynamical objects? After all, if one includes coherences, population crossings alone would not be sufficient to cause non-invertibility, as the coherences would also have to cross at the same time as the populations for two rows to be equivalent. Upon switching to the Argyres-Kelley projector in this unbiased regime, we indeed find a marked change in the profile of $\mathcal{R}(t)$ (see $\mc{R}(t)$ in Fig.~\ref{fig:CR_SB}, top panel). Only a single spike can still be observed before the onset of what is now many spikes at long times ($\sim9\rmm{--}10$~fs). Since we know that the physical source of these later spikes in $\mc{R}(t)$ arise from the approach to equilibrium of $\mc{C}(t)$, we need not consider those features further. However, the spike in $\mc{R}(t)$ at intermediate times ($\sim 5$~fs) is qualitatively different, $\mathcal{C}\neq\mathcal{C}_\rmm{eq}$. Critically, it obscures the region where one might expect $\mc{R}(t)$ to achieve its long-time limit $\mc{R}(\tau_R)$, and merits considering in more detail. 

To test the impact of the spike in $\mc{R}(t)$ at intermediate times, the bottom panel of Fig.~\ref{fig:CR_SB} shows the accuracy of the GME dynamics obtained when truncating $\mc{R}(t)$ at various trial cutoff times. This protocol corresponds to the assertion that the GME dynamics will be insensitive to the cutoff in $\mc{R}(t)$ beyond the point where it has reached its long time limit, $\mc{R}(\tau_R)$, and parallels the practice of cutting the memory kernel in the TC-GME at some particular time, $\tau_K$. Surprisingly, despite the presence of the spike, we find that $\mc{R}(t)$ reproduces the exact dynamics except for one unstable region just after the spike, between $4.6$ and $5.4$~fs. Long before this, the GME dynamics display monotonic convergence to the exact dynamics (black) with increasing trial cutoff time, $\tau$. This observation is recapitulated in the error plot (Fig.~\ref{fig:CR_SB}, inset), which considers the deviation of the GME dynamics from the reference exact dynamics as a function of cutoff time, $\tau$. While this error plot does not reveal the nature of the spike at intermediate time, it provides a criterion to choose the cutoff time, $\tau_R$, which may also be used in more complex scenarios. 

The unexpected robustness of the GME dynamics to the intermediate-time spike in $\mc{R}(t)$ in Fig.~\ref{fig:CR_SB} inspires two important questions: how can one understand the seemingly innocuous nature of the intermediate-time spike and, more worryingly, how can one successfully truncate $\mc{R}(t)$ if it does not, in fact, plateau? To answer these questions, we scrutinize the structure of $\mc{R}(t)$. For example, Fig.~\ref{fig:CR_SB}, top panel, suggests that there are $3$~different kinds of spikes in $\mc{R}(t)$: localized in time ($\mc{R}_{12}(t)$ and $\mc{R}_{42}(t)$), delocalized and symmetric ($\mc{R}_{32}(t)$), and delocalized and asymmetric ($\mc{R}_{22}(t)$). While one might expect that the highly localized spikes would not meaningfully impact the value of $\mc{R}$ at times beyond the immediate neighborhood of the spike, one cannot assume the same about the delocalized spikes, especially the asymmetric one. And yet, as noted earlier, the numerically extracted $\mc{R}(t)$ can recover the exact dynamics except when truncating $\mc{R}(t)$ in the unstable region in the middle of the spike, from $\tau \in [4.6, 5.4]$~fs. 

\begin{figure}[b]
\vspace{-18pt}
\begin{center}
    \resizebox{.5\textwidth}{!}{\includegraphics{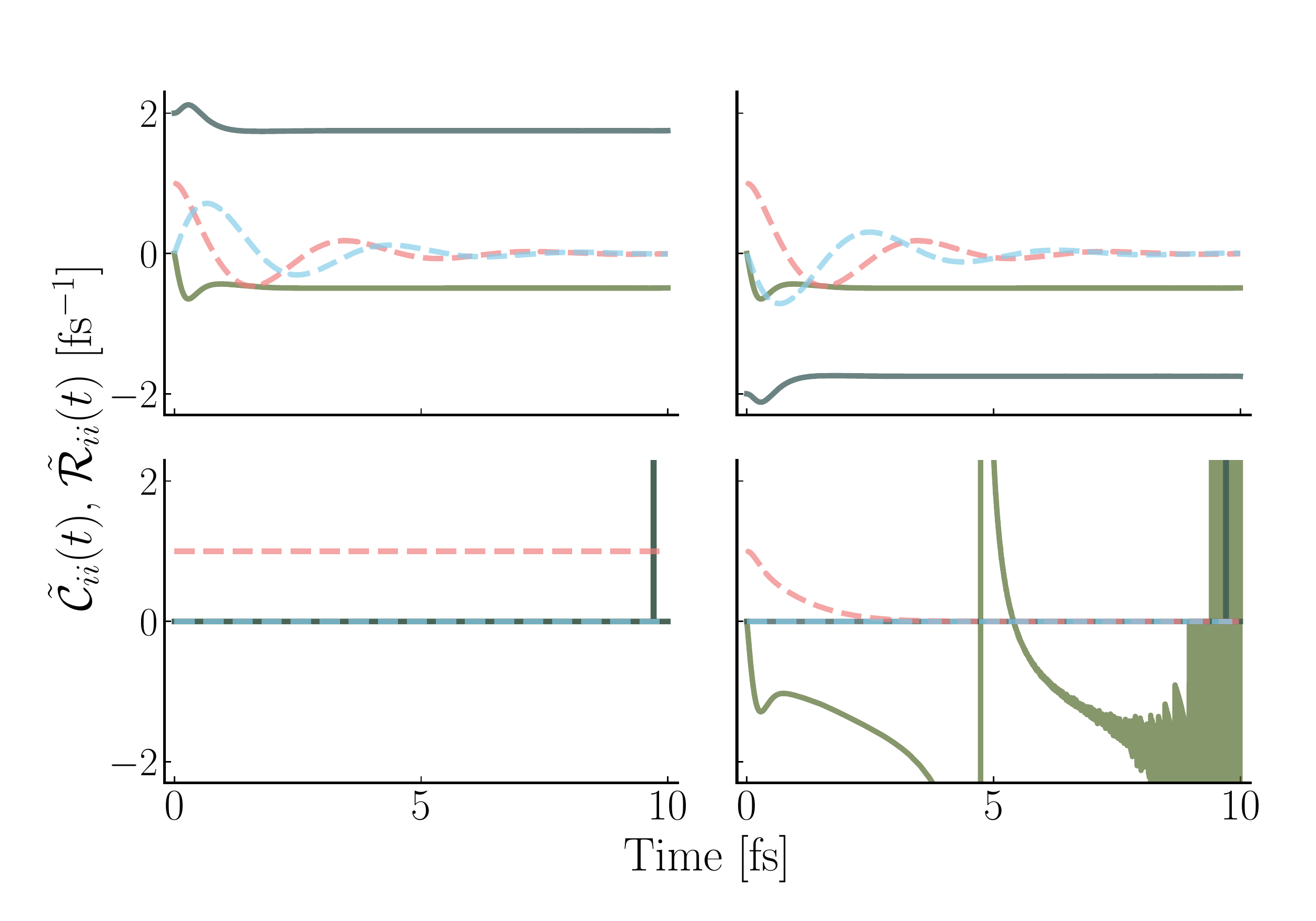}}\vspace{-2pt}
\end{center}
\vspace{-14pt}
\caption{\label{fig:CR_rot_SB} Diagonal elements of the rotated objects for the unbiased SB model with Argyres-Kelley projector shown in Fig.~\ref{fig:CR_SB}. $\tilde{\mc{R}}(t)$ is shown in as full green and teal lines, while $\tilde{\mc{C}}(t)$ is shown in dashed red and blue; real and imaginary parts, respectively. Only the final diagonal element exhibits a pole in the real part, and its dynamics are purely real and decaying.}
\end{figure}

To simplify the analysis of the poles in $\mc{R}(t)$ and their effects on the GME dynamics, we consider the distribution of nonzero elements in $\mc{R}(t)$. For example, $\mc{R}(t)$ has a banded structure, where only the second and third columns have a nonzero real part. This parallels what is seen in the TC-GME memory kernel (see Appendix~\ref{app:memorykernels}, Fig.~\ref{fig:memory_kernels_SB}). However, this apparent structure is only observed due to the choice of the site (diabatic) basis. Guided by the intuition that a diagonal representation would help elucidate the fundamental structure of $\mc{R}(t)$ and the influence of its spikes on the resulting GME dynamics, we rotate into the eigenbasis of the time-local generator at early times, i.e., $\mc{R}(0)$. This frame diagonalizes both $\mc{C}(t)$ and $\mc{R}(t)$ at early times but is not guaranteed to diagonalize both at later times. Yet, noting that for weak system-bath coupling cases, like the one considered in Figs.~\ref{fig:CR_SB} and \ref{fig:CR_rot_SB}, the time-local generator is not expected to change significantly as it equilibrates, the eigenbasis of $\mc{R}(0)$ should provide a predominantly diagonal representation for $\mc{C}(t)$ and $\mc{R}(t)$ over all times. Indeed, we observe that this basis effectively diagonalizes $\mc{R}(t)$ such that for our analysis we can neglect the off-diagonal elements (see Appendix~\ref{app:Rrotated} for further details) and examine only the diagonal elements, shown in Fig.~\ref{fig:CR_rot_SB}. The important progress this makes is that, in this simplified picture, an element $\tilde{R}_{ii}(t)$ interacts with its corresponding element in $\tilde{\mc{C}}_{ii}(t)$ \textit{only}. Now, we can inspect each diagonal element in isolation. 

In this new basis, the structure of $\tilde{\mc{R}}(t)$ (Fig.~\ref{fig:CR_rot_SB}) differs markedly from that of $\mc{R}(t)$ in the original basis (Fig.~\ref{fig:CR_SB}, top panel). First, $\tilde{\mc{R}}_{22}(t)$ (bottom left) is unity for all time, encoding the total conserved probability (i.e., the evolution of the trace of the density matrix subject to a normalized initial condition). This element can be truncated at any time. Second, $\tilde{\mc{R}}_{00}(t)$ and $\tilde{\mc{R}}_{11}(t)$ are conjugate elements that display damped oscillatory dynamics, similar to $\mc{C}(t)$ in the original basis, with non-zero imaginary components and negative real parts. These plateau at $\sim 4$~fs, in quantitative agreement with the $\tau_\mc{R}$ found in Fig.~\ref{fig:CR_SB}. The key panel is the last one. Here, $\tilde{\mc{C}}_{33}(t)$ is everywhere real, displaying critically damped behavior with no imaginary component. The real part of $\tilde{\mc{R}}_{33}(t)$ before the spike is negative, meaning that while $\lim_{t\rightarrow t_\rmm{pole}}\tilde{\mc{R}}_{33}(t)$ diverges, the corresponding $\lim_{t\rightarrow t_\rmm{pole}}\tilde{\mc{C}}_{33}(t) = 0$. Since $\tilde{\mc{C}}_{33}(t)$ reaches zero before the positive part of the spike in $\tilde{R}_{33}(t)$, the spike has no deleterious effect on the GME dynamics. Indeed, the time required for $\tilde{\mc{C}}_{33}(t)$ to reach zero is the same as $\tau_\mc{R}$. Importantly, this demonstrates that such spikes are innocuous as long as the elements of $\mc{C}(t)$ into which they multiply have decayed to zero. The caveat is that $\mc{R}(t)$ be real and negative for these elements.

If the requirement for a well-behaved, yet singular, $\tilde{\mc{R}}(t)$ is for the corresponding elements of $\tilde{\mc{C}}(t)$ to decay sufficiently fast to zero, then our analysis explains the difficulties previously seen in the TCL-GME obtained with the populations-only projector \cite{LiuShi2018} and in the limit of fast coherence relaxation when using the full projector \cite{Kropf2016}. For the populations-only projector, all elements are positive and real. In the rotated basis, one of the diagonal elements remains at one for all time (conservation of probability), whereas the second diagonal decays to zero with the equilibration of the system. This follows from inspecting the trace of $\mc{C}(t)$, which goes from $2$ at initial times to $1$ at long times. Since the trace is invariant to unitary rotations, in the rotated basis the second diagonal must start at $1$ and decay to $0$ at long times. Clearly, therefore, the dynamics cannot decay to zero before the onset of pre-equilibrium spikes in the rotated frame. Hence, the regular set of poles that appear from population crossings at periods of the generalized Rabi frequency unavoidably break the dynamics. 

The above analysis provides important physical insight into the spike structure of $\mc{R}(t)$ and the potential sensitivity of the GME dynamics to the spikes. For example, the fact that the spikes that appear in the banded structure of $\mc{R}(t)$ in the original basis do not negatively affect the GME dynamics makes sense when one considers that these elements multiply into the rows representing initial coherences, which become conjugate at long times. It furthermore explains why the dynamics are unstable in the region around $4.8$~fs. As a floating point number, $\tilde{\mc{C}}_{33}(t)\neq 0$, which allows the repeated application of a large, positive $\tilde{\mc{R}}_{33}(t\gtrapprox t_\rmm{pole})$ to cause it to diverge. Indeed, one finds that the more one increases $\tau$, the longer the dynamics take to diverge, exactly because $\tilde{\mc{R}}(t)$ is rapidly decreasing. Eventually, at $\tau \approx 5.4$~fs, $\tilde{\mc{R}}_{33}(t)$ becomes negative, and the dynamics are once again stable. The short period of numerically divergent $\tilde{\mc{R}}_{33}(t)$ that is included before cutoff for such values of $\tau$ is not sufficient in duration to cause the dynamics to become unstable. And yet, as this analysis demonstrates, this delocalized asymmetric spike is important to the dynamics. 

The robustness of the TCL-GME will therefore depend on a competition between the timescales of two processes:
\begin{enumerate}
    \item The onset of spikes: the accidental equivalence of measurements arising from two different initial conditions \cite{Maldonado-Mundo2012}. For example, we see that under a populations-only projector this happens every time the populations cross, although this requires the problem to be symmetric (unbiased, $\epsilon=0$). In contrast, if the coherences are included with an Argyres-Kelley projector, only one (pre-equilibrium) spike is observed for the same parameters.
    
    \item $\tau_\mc{R}$: the maximum of a) the time it takes for $\tilde{\mc{R}}(t)$ to plateau in the absence of spikes---see Fig.~\ref{fig:CR_rot_SB}, top panels---and b) the time it takes for $\tilde{\mc{C}}(t)$ to decay to zero even when the relevant element(s) of $\tilde{\mc{R}}(t)$ are still changing, and possibly even diverging (see Fig.~\ref{fig:CR_rot_SB}, bottom right panel).\footnote{We note that in our analysis surrounding Fig.~\ref{fig:CR_rot_SB}, a) and b) are the same.}
\end{enumerate}
If the first timescale is faster, an exact time-local description does not exist. The greater the discrepancy, the worse the minimum-error choice of $\tau$ will become. In contrast, if the latter (decay) timescale is fastest, then the time-local generator should be able to recover the dynamics for all time. This analysis therefore suggests that the success and feasibility of a time-local description of non-Markovian dynamics will \textit{depend on both the Hamiltonian parameters and the choice of projector.}

\begin{figure}[t]
\begin{center}
    \resizebox{.5\textwidth}{!}{\includegraphics{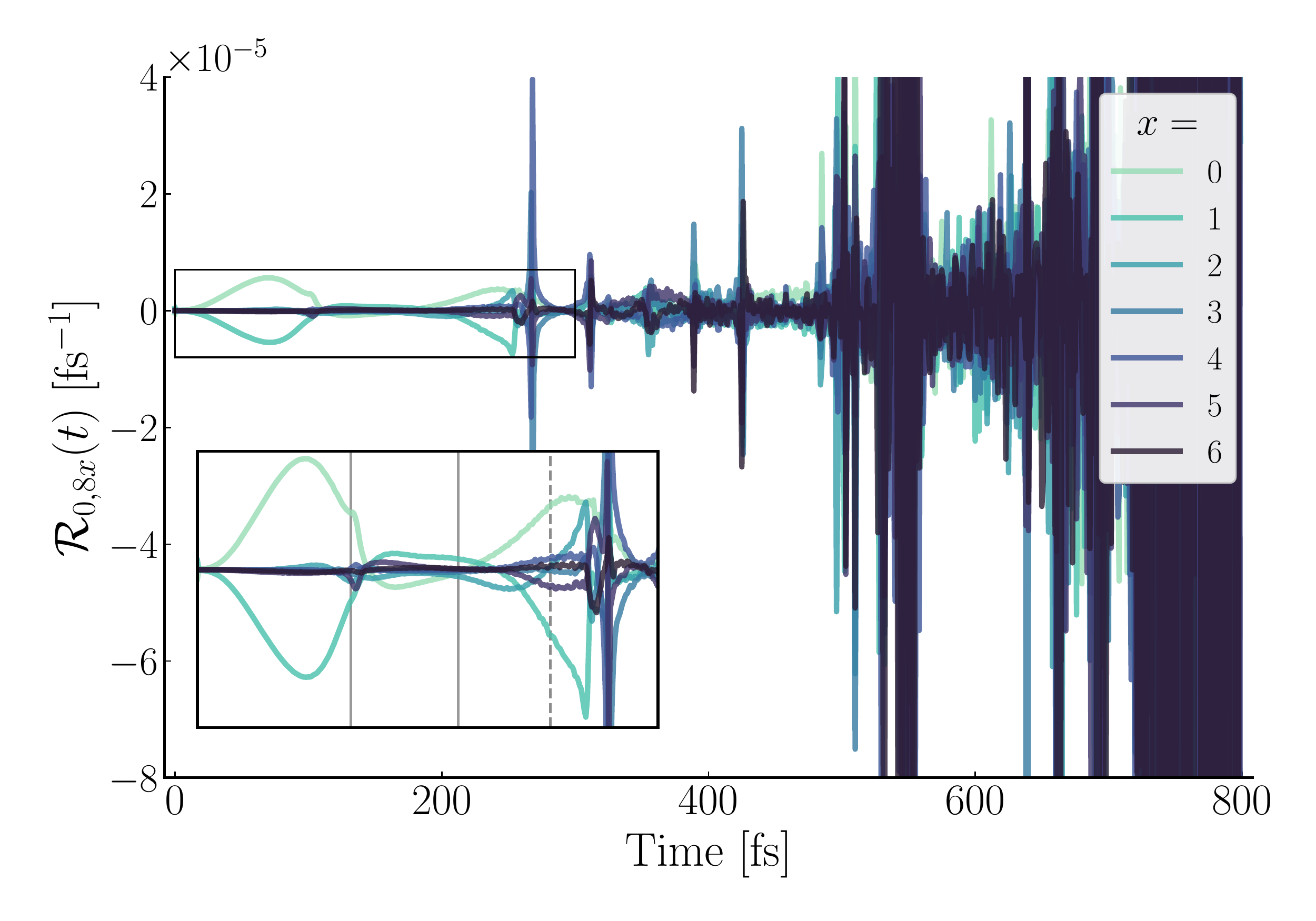}}\vspace{-2pt}
\end{center}
\vspace{-18pt}
\caption{\label{fig:R_FMO} FMO model $\mc{R}(t)$ for the population elements of its first row where $\tau_B=166$~fs. The total matrix is $49 \times 49$. \textbf{Inset}: zoomed in view of the region before onset of the first spikes. Vertical grey lines correspond to the cutoffs discussed in the next figure, while the dashed grey line is the best choice of cutoff (for all matrix elements, including those not displayed here).}
\end{figure}

\subsection{Application to a multi-state model} 

We now turn to the more challenging one-exciton dynamics of the 7-site Fenna-Matthews-Olson (FMO) model. The thoroughly studied FMO model serves as a test bed for the virtues and flaws of various quantum dynamics approaches, making it a logical candidate to test our approach to the TCL-GME. In particular, we focus on two parameter regimes of the FMO complex, corresponding to a slow bath with an average decorrelation time of $\tau_B = 166$~fs and a fast bath with $\tau_B = 50$~fs. These parameter regimes allow us to further disentangle the influence of the Hamiltonian parameters and their interplay with the projection operator that lead to distinct performances for the TCL-GME. 

\begin{figure}[t]
\begin{center}
\vspace{-8pt}
    \resizebox{.5\textwidth}{!}{\includegraphics{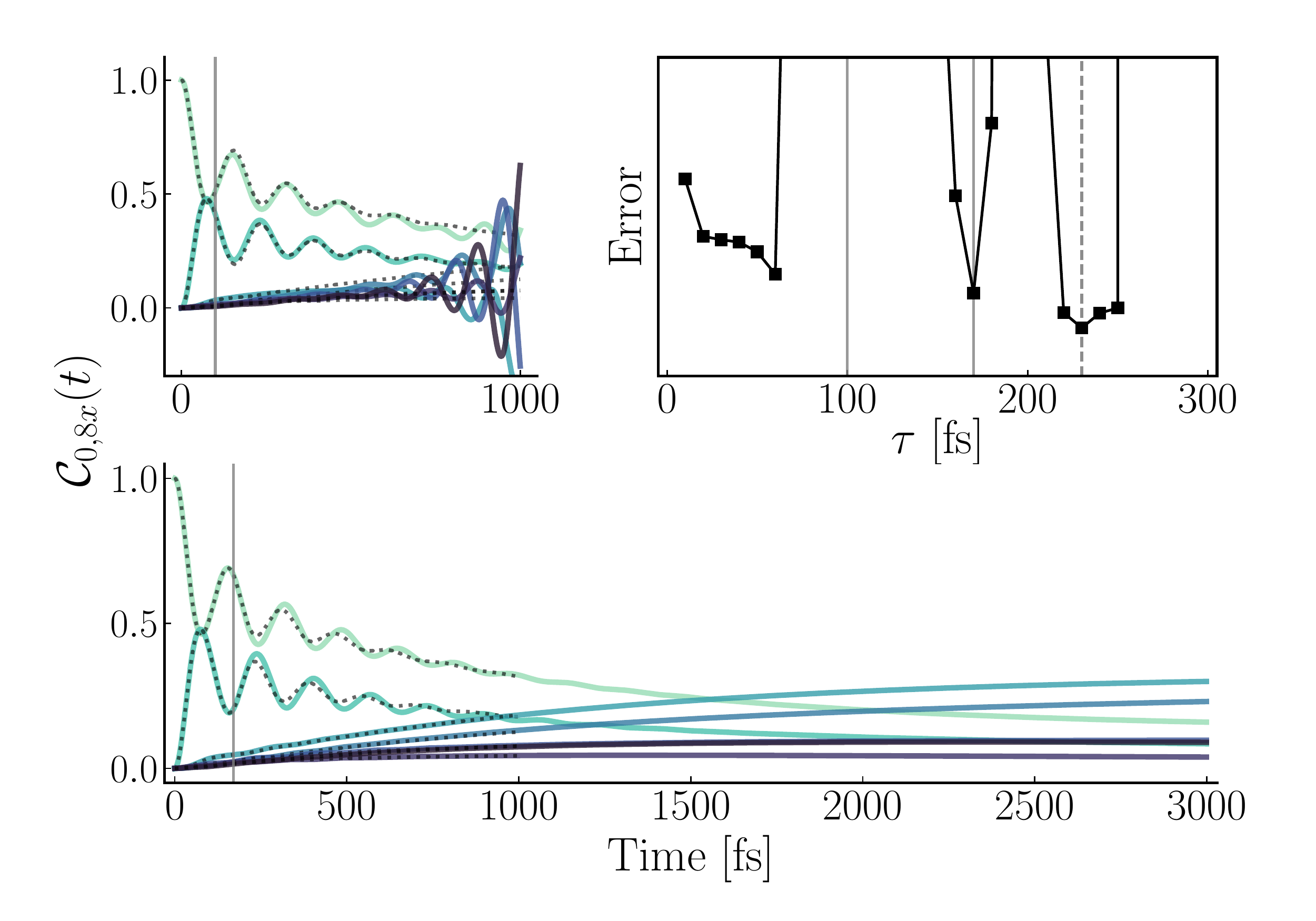}}\vspace{-6pt}
\end{center}
\vspace{-12pt}
\caption{\label{fig:Cs_FMO} $\mc{C}(t)$ elements corresponding to $\mc{R}(t)$ in Fig~\ref{fig:R_FMO}; the population correlations given initialization in the first site. The HEOM dynamics are shown as grey dots. Vertical grey lines are the cutoffs, $\tau$. \textbf{Top Left}: $\tau=100$~fs, where $\mc{R}(t)$ may have temporarily plateaued. \textbf{Bottom}: $\tau=170$~fs, showing stability even at long times. \textbf{Top Right}: Plot of the error versus $\tau$ showing that there is another region of stability around $\tau=230$~fs (dashed grey line), after which the dynamics diverge permanently.}
\end{figure}

Here we focus on the full density matrix dynamics for a widely used parameterization of the FMO model \cite{Adolphs2006} with a slow nuclear bath decorrelation time of $\tau_B = 166$~fs (for simulation details, see Appendix~\ref{app:HEOMsettings}). Figure~\ref{fig:R_FMO} shows that $\mc{R}(t)$ for this parameter regime starts to display spikes by $\sim250$~fs. If one uses the full $\mc{R}(t)$, one can capture the exact dynamics of $\mc{C}(t)$ over the first $\sim$800~fs, suggesting that it is only the spikes after this time that lead to inaccurate GME dynamics. This reiterates that the TCL-GME is remarkably robust to some spikes. Given the congestion of spikes at later times, can we still use the error plot to determine a good choice for $\tau_R$? Such an error plot, in top right of Fig.~\ref{fig:Cs_FMO}, demonstrates that the GME dynamics will be well-behaved at cutoff times around 170~fs and 230~fs. Figure~\ref{fig:Cs_FMO} illustrates the GME dynamics obtained when one truncates $\mc{R}(t)$ at $\tau = 100$~fs (top left) and $\tau = 170$~fs (bottom). The agreement for $\tau = 100$~fs is perfect for $t \leq \tau$, acceptable until $t \approx 600$ fs, but deviates unphysically at longer times. In contrast, at $\tau = 170$ fs, the agreement remains satisfactory over the entire time the numerically exact dynamics are available ($t \leq 1000$ fs) and continues to behave as expected over indefinitely longer timescales. And yet, even with a truncation time of $\tau=170$~fs, the resulting TCL-GME dynamics do not agree perfectly with the exact dynamics. In fact, the memory kernel in the TC-GME treatment of the same problem decays on a timescale of $\tau_\mc{K} \approx 600$~fs (see Appendix~\ref{app:memorykernels}, Fig.~\ref{fig:memory_kernels_FMO}). 

Unfortunately, truncating $\mc{R}(t)$ at a similar timescale to the TC-GME leads to ill-behaved long-time dynamics, suggesting that some problems displaying non-Markovian dynamics cannot be exactly expressed in a time-local fashion. Nonetheless, these results suggest that, if one is ready to accept a little inaccuracy in the resulting dynamics for particularly challenging cases, the TCL-GME provides the most compact and complete description of the non-Markovian dynamics of $\mc{C}(t)$. Indeed, here the relaxation dynamics of the one-exciton manifold of the FMO complex take $\sim 10$~ps, whereas the time required to capture $\mc{R}(t)$ so as to reproduce the dynamics to the accuracy of Fig.~\ref{fig:Cs_FMO} is only the first 170~fs, \textit{two orders of magnitude less than the equilibration time.} 

Despite the reasonable agreement, these time-local dynamics fall short of the near-exact accuracy observed in Fig.~\ref{fig:biased-SB-model-good-example}. Given the relatively significant difficulties experienced in describing this parameter regime of FMO with the TCL-GME, can we employ our observation of a competition of timescales (between spike onset versus decay) to shed light on the feasibility of a time-local description? From this perspective, one would predict that the spikes in $\tilde{\mc{R}}(t)$ appear before some purely real and dissipative component(s) of the rotated dynamics decay in the eigenbasis of $\mc{R}(0)$. Our calculations show that this prediction about $\tilde{\mc{C}}(t)$ is indeed borne out (See Appendix~\ref{app:Rrotated}, Fig.~\ref{fig:rotated_baths}) in this larger ($49$~diagonal element) space.

\begin{figure}[t]
\begin{center}
\vspace{-12pt}
     \resizebox{.5\textwidth}{!}{\includegraphics{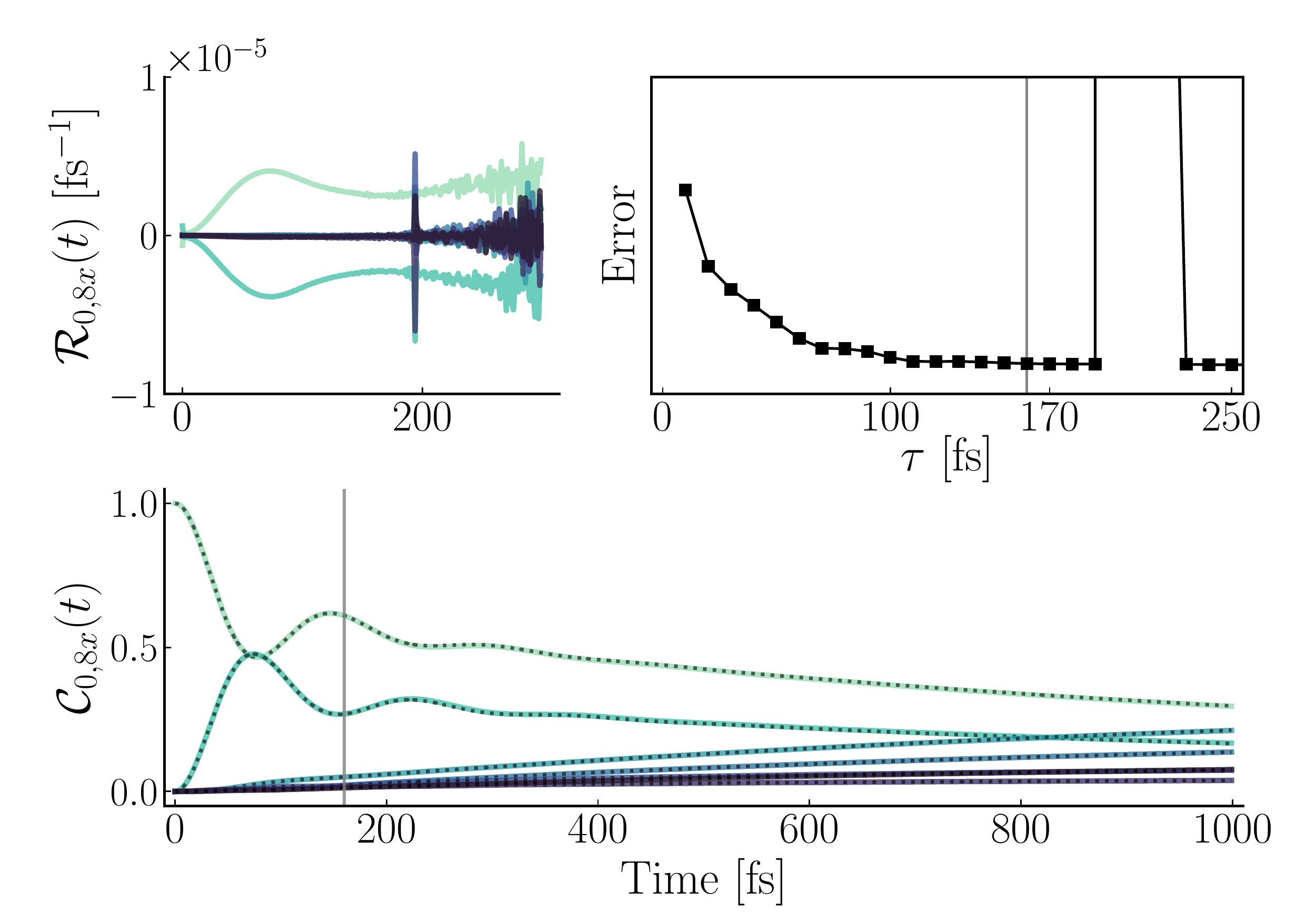}}\vspace{-4pt}
\end{center}
\vspace{-10pt}
\caption{\label{fig:CR_fast_FMO} Faster bath, all vertical grey lines showing $\tau = 160$~fs. \textbf{Top Left}: $\mc{R}(t)$ equivalent to the inset of Fig.~\ref{fig:R_FMO}. Note that the position of the first spike has moved to an earlier time, but that $\mc{R}(t)$ has now visibly reached a plateau before this time. \textbf{Bottom}: Plot of GME dynamics (colors) compared to HEOM dynamics (grey), equivalent to Fig.~\ref{fig:Cs_FMO}. \textbf{Top Right}: The average error of the GME dynamics after $\tau$, determining that $\tau_\mc{R} = 160 \pm 5$~fs$^{-1}$. Clearly, around $100$~fs$^{-1}$ the results are fairly insensitive to $\tau$. }
\end{figure}

It follows, then, that reducing the decay timescale would improve the performance of the TCL-GME. To test this, we consider the one-exciton dynamics of the FMO complex with an alternative fast nuclear bath decorrelation time of $\tau_B = 50$~fs \cite{Ishizaki2009b, Cho2005a}. While increasing the speed of the bath may also accelerate the onset of spikes, we test our hypothesis nevertheless, and present the results in Fig.~\ref{fig:CR_fast_FMO}. As is evident from the the top~left panel, the onset of spikes in $\mc{R}(t)$ now occurs at $\sim180$~fs, a little after what appears to be a plateau in $\mc{R}(t)$. The error plot (top right of Fig.~\ref{fig:CR_fast_FMO}) confirms this claim, showing monotonically decreasing error in the GME dynamics as a function of increasing cutoff until before the spike at $180$~fs. The GME dynamics obtained with $\tau_R = 160$~fs shown in the bottom panel of Fig.~\ref{fig:CR_fast_FMO} recover the exact dynamics without incident. While our approach is nonperturbative, this accords with the traditional, perturbative understanding that the time-local approach performs better in regimes of weaker coupling \cite{BreuerPetruccione}.

\section{Discrete-time TCL-GME} 
\label{sec:discrete-time-tcgme}

Thus far, we have focused on a continuous-time formulation of the TCL-GME that would be compatible with both numerically exact quantum dynamics as well as classical mechanical approaches. Yet, there are many approaches that can only afford limited resolution in time or which suffer from noisy data that render numerical derivatives unstable \cite{Muhlbacher2008, Chatterjee2019, Gull2011, Chen2017}. For these cases, a discrete-time analogue to the TTM in the time-nonlocal formulation would resolve this difficulty. To construct the discrete-time analogue of the TCL-GME, we formally integrate Eq.~(\ref{eq:TCL-GME}) to obtain
\begin{equation}\label{eq:integrated-discrete-time-TCL-GME}
    \mc{C}(t+\delta t) = \mc{U}(t+\delta t, t) \mc{C}(t).
\end{equation}
where $\mc{U}(t+\delta t, t) = \exp_{\rightarrow}\big[\int_{t}^{t+\delta t} ds\ \mc{R}(s)\big]$ and the $\rightarrow$ subscript denotes time-ordering of the exponential. Similar to our direct inversion of the TCL-GME in Eq.~(\ref{eq:TCL-GME}), for the continuous-time version of our approach we isolate the non-Markovian propagator in Eq.~(\ref{eq:integrated-discrete-time-TCL-GME}),
\begin{equation}\label{eq:isolated-non-Markovian-propagator}
    \mc{U}(t+\delta t, t) = \mc{C}(t+\delta t)[\mc{C}(t)]^{-1},
\end{equation}
which becomes a simple function of the time difference, $\delta t$, for $t \geq \tau_R$,
$\mc{U}(t+\delta t, t) = \exp\big[ \mc{R}(\tau_R)\delta t]$. 

\begin{figure}[!h]
\vspace{-10pt}
\begin{center}
    \resizebox{.5\textwidth}{!}{\includegraphics{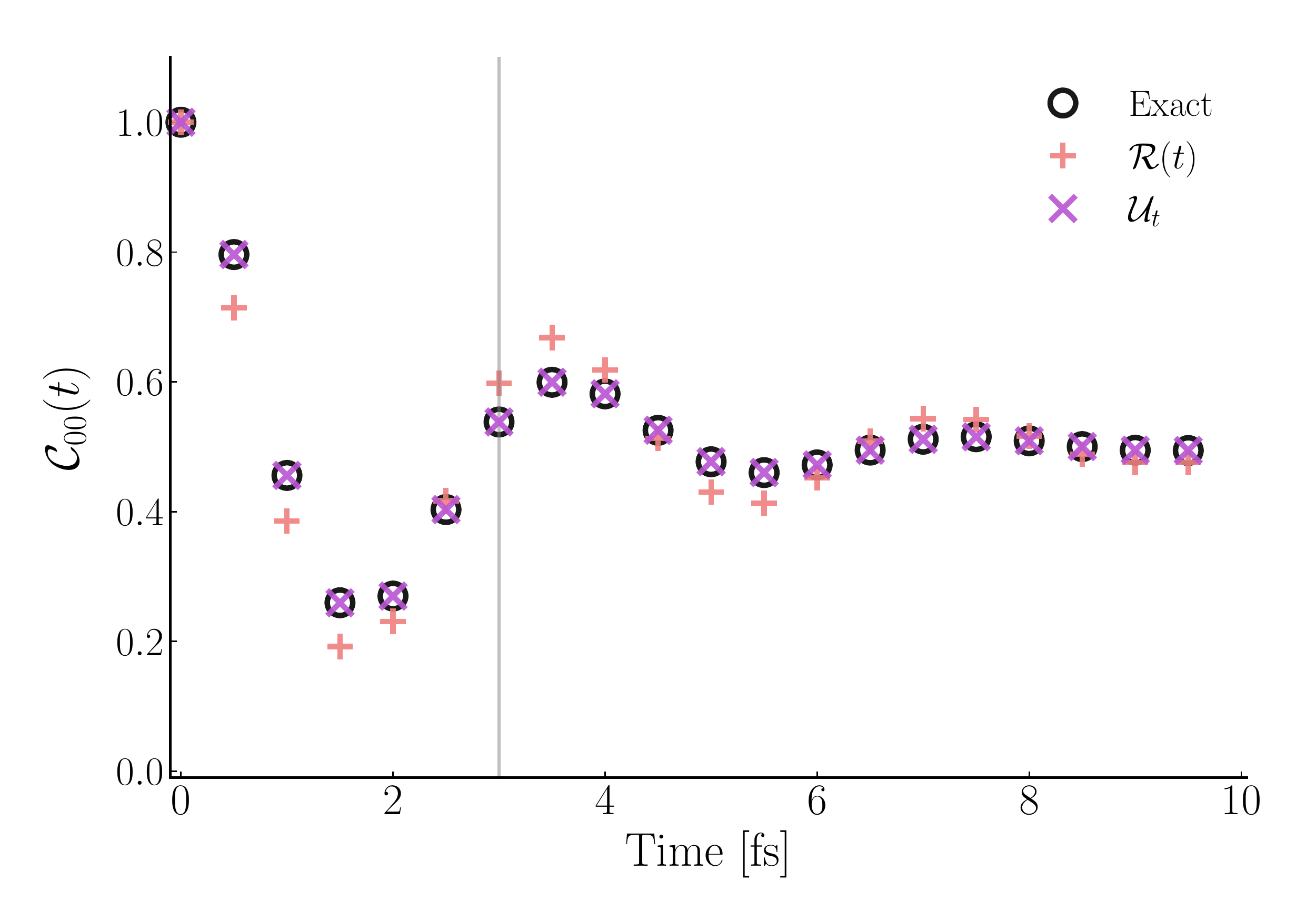}}\vspace{-2pt}
\end{center}
\vspace{-14pt}
\caption{\label{fig:pathintegral_SB} Subsampled HEOM data used to represent discrete-time data arising from a method such as QUAPI. This sparsity amounts to subsampling 1/500 points from the HEOM, but the data can be made arbitrarily sparse for the $\mc{U}_t$ method. Using $\mc{R}(t)$ leads to visible error even before any cutoff is made. $U_t$ however has no error, and can still be cut off at the vertical grey line.}
\end{figure}

We demonstrate the feasibility of this approach by decimating the resolution of the exact dynamics we used from the SB model of Fig.~\ref{fig:CR_SB}. Figure~\ref{fig:pathintegral_SB} shows the perfect agreement between our discrete-time TCL-GME and the low-resolution version of the numerically exact population dynamics shown previously. In contrast, employing the continuous-time version of the TCL-GME to construct a low-resolution $\mc{R}(t)$ gives poor agreement with the reference data when used to propagate $\mc{C}(t)$ using Eq.~(\ref{eq:TCL-GME}). Simply, the integration step is not valid. In fact, it follows that by obviating the integration step this discrete-time rewriting can actually tolerate poles that the continuous-time version could not. In addition, the discrete-time version allows us to resolve the previously observed problems in the divergent time-local generator for the populations-only projector for the unbiased SB model in Fig.~\ref{fig:popsOnly_SB} (see Appendix~\ref{app:SVD_discussion}).  These demonstrations establish the applicability of our discrete-time TCL-GME approach for exact data with low temporal resolution, reiterating that the TCL-GME provides the most compact and complete description of the non-Markovian dynamics.

\section{Conclusion} 
\vspace{-4pt}

Here we have shown how one can combine the TCL-GME, both in continuous and discrete time, with numerically exact approaches that capture the classical or quantum dynamics of complex systems to obtain a compact and complete representation of the non-Markovian dynamics of reduced observables. In particular, we have provided a simple scheme for constructing the continuous time-local generator $\mc{R}(t)$ using only the reference dynamics, $\mc{C}(t)$, and its numerical time derivative, $\dot{\mc{C}}(t)$. In addition, we have provided a straightforward approach to obtain the discrete-time propagator, $\mc{U}(t+ \delta t, t)$, requiring only the reference dynamics and no time derivatives, which is applicable to cases where the temporal resolution of the dynamics is not sufficient to accurately capture the time-derivative $\dot{\mc{C}}(t)$ required to construct the continuous-time $\mc{R}(t)$. This discrete-time approach has recently been employed and extended to tame noisy reference dynamics and capture the long-time conformational dynamics of protein folding \cite{Dominic2022}. 

We have further demonstrated how and when the time-local approach is able to recapitulate the exact dynamics from which it is built and when the presence of divergences or near-divergences (spikes) in the time-local generator allow only an approximate treatment. We have analyzed the origin and severity of such spikes and demonstrated that, unlike previously thought, the time-convolutionless approach is robust to a subset of these spikes under conditions that we have established. Interestingly, we find that even in cases where the time-local generator does not exist (i.e., where identifying a well defined $\tau_R$ is not possible), early truncation of $\mc{R}(t)$ still yields approximate dynamics that are well behaved (i.e., Fig.~\ref{fig:Cs_FMO}). Moreover, we have shown that when the time-local generator exists, the TCL-GME provides a more compact, intuitive, and complete description of the non-Markovian projected dynamics than the TC-GME, which has already been shown to offer a more data-efficient means to encode projected dynamics than effective Markovian theories like Markov state models \cite{CaoMontoya2020}. 

Thus, our present work demonstrates how one can exploit the TCL-GME to simply and effectively extend the reach of dynamical approaches where long-time dynamics are computationally expensive or where statistical error inhibits their calculation. The insights presented here outline a path for future applications of the time-local formalism and highlight fundamental questions about the existence of a time-local description of non-Markovian dynamics and its ability to provide the most compact and complete representation of such dynamics.

\vspace{-10pt}
\section{\label{sec:acknow} Acknowledgments}
\vspace{-6pt}
We thank William Pfalzgraff for reading the manuscript and useful comments and Joel Eaves and Sandeep Sharma for helpful discussions. We acknowledge start-up funds from the University of Colorado Boulder.

\appendix
\section{Computational Details}\label{app:compdeets}
\vspace{-10pt}
\subsection{Hierarchical Equations of Motion}\label{app:HEOMsettings}
We performed all HEOM calculations using the open-source pyrho code \cite{Berkelbach2020}. For the systems in the main text, the parameter regimes are, in order of appearance:
\begin{enumerate}
    \item Biased spin-boson models of Fig.~\ref{fig:biased-SB-model-good-example}: $\epsilon = 1$, $k_\rmm{B}T = 1/5$, $\omega_c = 1$, $\lambda = 0.05$; in energy units where $\Delta = 1$; $K=9$ and $L=3$ with $dt=0.001$; with the superohmic spectral density, i.e., $J(\omega) = \frac{\pi\lambda}{3\omega_c^3}\omega^3 e^{-\omega/\omega_c}$.
    
    \item Unbiased spin-boson models of Figs.~\ref{fig:popsOnly_SB}, \ref{fig:CR_SB}, \ref{fig:CR_rot_SB} and \ref{fig:pathintegral_SB}: $\epsilon = 0$, $k_\rmm{B}T = 10$, $\omega_c = 2$, $\lambda = 0.1$, in energy units where $\Delta$ = 1; $K=0$ and $L=3$ with $dt=0.001$; with a superohmic spectral density.
    
    \item Frenkel exciton model of Fig~\ref{fig:R_FMO} and \ref{fig:Cs_FMO}: $T = 300~\rmm{K}$, $\lambda=35$ cm$^{-1}$, $\tau=166$~fs such that $\omega=1/\tau$; $K=0$ and $L=4$ with $dt=1$ fs. Each site in the Frenkel exciton model is coupled to a local bath where the coupling is described by an Ohmic-Lorentz spectral density, $J(\omega) = \frac{2 \lambda \omega}{\omega^2 + \omega_c^2}$. 
    
    \item Frenkel exciton model of Fig~7: as above but with $\tau=50$~fs.
\end{enumerate}

For completeness, the forms of the SB and Frenkel exciton models used here can be written as follows, 
\begin{equation}
    H = H_{\rm S} + H_{\rm B} + H_{\rm SB}.
\end{equation}
These three terms correspond to the system, bath, and system-bath Hamiltonians. 

For both the SB and Frenkel exciton models, the system Hamiltonian has the same form, 
\vspace{-4pt}
\begin{equation}
    H_{\rm S} = \sum_{j,k}^{{\rm N_{Hil}}} h_{j,k} \ket{j}\bra{k}.
\end{equation}
In the SB model,
\vspace{-10pt}
\begin{equation*}
    \mathbf{h} = \begin{bmatrix}
\varepsilon & \Delta \\
\Delta  & -\varepsilon
\end{bmatrix}.
\end{equation*}
For the Frenkel exciton model of the FMO complex,
\begin{equation*}
\mathbf{h} = \begin{bmatrix}
12410 & -87.7 & 5.5 & -5.9 & 6.7 & -13.7 & -9.9\\
-87.7 & 12530 &  30.8 &   8.2 &   0.7 &  11.8 &  4.3\\
5.5 &  30.8 & 12210 & -53.5 &  -2.2 &  -9.6 &   6.0\\
-5.9 &   8.2 & -53.5 & 12320 & -70.7 & -17.0 & -63.3\\
6.7 &   0.7 &  -2.2 & -70.7 & 12480 &  81.1 &  -1.3\\
-13.7 &  11.8 &  -9.6 & -17.0 &  81.1 & 12630 &  39.7\\
-9.9 &   4.3 &   6.0 & -63.3 &  -1.3 &  39.7 & 12440
\end{bmatrix}
\end{equation*}
in units of $\rmm{cm}^{-1}$.

For both the SB and Frenkel exciton models, the bath is composed of independent harmonic oscillators. The main difference lies in the fact that for the SB model, there is one antisymmetrically coupled bath, whereas in the Frenkel exciton model, each site is connected to its local bath. Thus, for the SB model
\vspace{-4pt}
\begin{equation}
    H_{\rm B} = \frac{1}{2}\sum_{n} [\hat{p}_n^2 + \omega_n \hat{q}_n^2],
\end{equation}
and
\vspace{-10pt}
\begin{equation}
    H_{\rm SB} = \sigma_z \sum_{n} c_n \hat{q}_n,
\end{equation}
where $\sigma_z$ is the $z$ Pauli matrix, $\hat{q}_n$ and $\hat{p}_n$ are the mass-weighted position and momentum operators for the $n^\rmm{th}$ harmonic oscillator in the bath, $\omega_n$ is the frequency of the $n^\rmm{th}$ oscillator and $c_n$ its coupling constant to the spin. The couplings are given by the spectral density of the system, 
\vspace{-10pt}
\begin{equation}
    J(\omega) = \frac{\pi}{2}\sum_{n} \frac{c_n^2}{\omega_{n}}\delta(\omega - \omega_n).
\end{equation}

For the Frenkel exciton model,
\vspace{-4pt}
\begin{equation}
    H_{\rm B} = \sum_{k}^{{\rm N_{Hil}}} H_{\rm B}^{(k)},
\end{equation}
and 
\vspace{-10pt}
\begin{equation}
    H_{\rm SB} = \sum_{k}^{{\rm N_{Hil}}} \ket{k}\bra{k}\sum_{n} c_{k,n} \hat{q}_{k,n},
\end{equation}
where now each harmonic oscillator contains two labels: the first, $k$, labels the electronic site to which it belongs, and the second, $n$, identifies that harmonic oscillator within the local bath. Similarly, the coupling constants of each site to its local bath, $c_{k,n}$, are given by the local spectral density,
\begin{equation}
    J_k(\omega) = \frac{\pi}{2}\sum_{n} \frac{c_{k,n}^2}{\omega_{k,n}}\delta(\omega - \omega_{k,n}).
\end{equation}

\vspace{-16pt}
\subsection{Time-local Generalized Master Equation}\label{app:Rconstruct}
To obtain the time-local generator we first construct the Liouville matrix $\mc{C}(t)$ in Eq.~(\ref{eq:Liouville-dynamical-matrix}). Depending on the choice of projector, $\mc{C}(t)$ contains either all or a subset of the reduced (electronic) density matrix elements subject to all multiplicative initial conditions as a function of time. 

To construct $\mc{C}(t)$ in the case of the Argyres-Kelly projector, which yields the entire density matrix, one needs to run $(N_{\rm Hil}^{\rm elec})^2$ HEOM calculations, where $N_{\rm Hil}^{\rm elec}$ is the number of distinct electronic states, $\ket{j}$, that span the electronic subsystem Hamiltonian. These calculations correspond to the measurement of all distinct electronic measurements, $|i\rangle\langle j|$ for $i,j \in \{1, ..., N_{\rm Hil}^{\rm elec}\}$ subject to all distinct multiplicative initial conditions, $|k\rangle\langle l|\rho_B$ for $k,l \in \{1, ..., N_{\rm Hil}^{\rm elec}\}$. Each HEOM calculation of the a particular reduced density matrix subject to a particular initial condition yields a $N_{\rm Hil}^{\rm elec} \times N_{\rm Hil}^{\rm elec}$ matrix, which one reorders into a $N_{\rm Hil}^{\rm elec} \times 1$ vector. This vector becomes the $(N_{\rm Hil}i+j)^\rmm{th}$ row of $\mc{C}(t)$. For the SB model, where $N_{\rm Hil}^{\rm elec} = 2$, this results in $\mc{C}(t)$ being a $4 \times 4$ time-dependent matrix, whereas for the Frenkel exciton, where $N_{\rm Hil}^{\rm elec} = 7$, this results in $\mc{C}(t)$ being a $49 \times 49$ time-dependent matrix. In the case of the populations-only projector, we simply omit the elements $i\neq j$ and $k \neq l$ terms when constructing $\mc{C}(t)$. This implies that the resulting $\mc{C}(t)$ for the populations-only projector has a dimensionality of $N_{\rm Hil}^{\rm elec} \times N_{\rm Hil}^{\rm elec}$. 

From the numerically exact $\mc{C}(t)$, we construct $\mc{R}(t)$ using Eq.~(\ref{eq:R-from-inverse}). That is, we take the finite difference derivative of $\mc{C}(t)$ and right-multiply it by the inverse of $\mc{C}(t)$ (at each time step). Although $\mc{C}^{-1}(t)$ sometimes diverges analytically, at this resolution we do not encounter overflow errors in the numerics, only large values in $\mc{R}(t)$.

In this work, when predicting the dynamics using $\mc{R}(t)$ under some cutoff, we simply compute $\dot{\mc{C}}(t)$ at each time step from Eq.~(\ref{eq:TCL-GME}) by left-multiplying by $\mc{R}(t)$, and evolve the $\mc{C}(t)$ matrix from its initial condition (the identity matrix) using Heun's method.

For the discrete-time version in Eqs.~(\ref{eq:integrated-discrete-time-TCL-GME}) and (\ref{eq:isolated-non-Markovian-propagator}), we obtain $\mc{U}(t+\delta t,t)$ equivalently from right-multiplying $\mc{C}(t+\delta t)$ --- rather than its derivative --- by $\mc{C}^{-1}(t)$. Therefore numerical time-derivatives are unnecessary and one can trivially evolve $\mc{C}(t)$ by left-multiplying it by $\mc{U}(t+\delta t,t)$ as Eq.~(\ref{eq:integrated-discrete-time-TCL-GME}) suggests.

\section{Population-only Integration}\label{app:SVD_discussion}
The majority of `spikes' encountered in this work are not true poles, in the sense that they do not cause overflow errors in the numerics. However, the first pole in the populations-only projector dynamics displayed in Fig.~\ref{fig:popsOnly_SB} is sufficiently divergent to cause the error in the integration of the equation of motion encapsulated in that figure. As discussed in the main text, for this highly symmetric problem the spikes occur whenever the populations cross. Uniquely, all spikes therefore belong to the null space of equilibrium vectors. It stands to reason that in this case we could analytically manipulate the structure to eschew integrating the poles altogether\cite{Cohen2013b, Amati2022}. 

Here, we take a purely numerical, practical view of the problem to maintain maximum transferability to other regimes and models where the manipulations are more demanding and/or the poles do not all belong to the same space. The most immediate manipulation to perform is a pseudo-inverse. By writing the correlation matrix as a diagonal matrix rotated by two complex unitary matrices 
\begin{equation}
    \mathcal{C}(t) = \mathcal{W}(t)\Sigma(t)\mathcal{V}^\dagger(t),
\end{equation}
i.e., a singular-value decomposition (SVD) at each time step, we can remove the singularities from the inverse by only taking the reciprocal of the non-zero entries of $\Sigma(t)$ to yield the pseudo-inverse singular matrix
\begin{equation}
    \sigma_{ij}^{-1}(t) = 
    \begin{cases}
        0 \textrm{ if } \Sigma_{ij}=0,\\
        1/\Sigma_{ij}(t),\textrm{otherwise.}
    \end{cases}
\end{equation}
Then, the pseudo-inverse of the correlation matrix is
\begin{equation}\label{eq:SVD}
    \mathcal{C}^{-1}(t) = \mathcal{V}(t)\sigma^{-1}(t)\mathcal{W}^\dagger(t).
\end{equation}
Numerically, we define a threshold of $10^{-5}$ to be small enough to be considered zero. In the case of Fig.~\ref{fig:popsOnly_SB}, only the first and last poles actually meet this criterion, which explains why only this first pole is `singular enough' to cause clear divergence. 

\begin{figure}[!t]
\vspace{-18pt}
\begin{center}
    \resizebox{.5\textwidth}{!}{\includegraphics{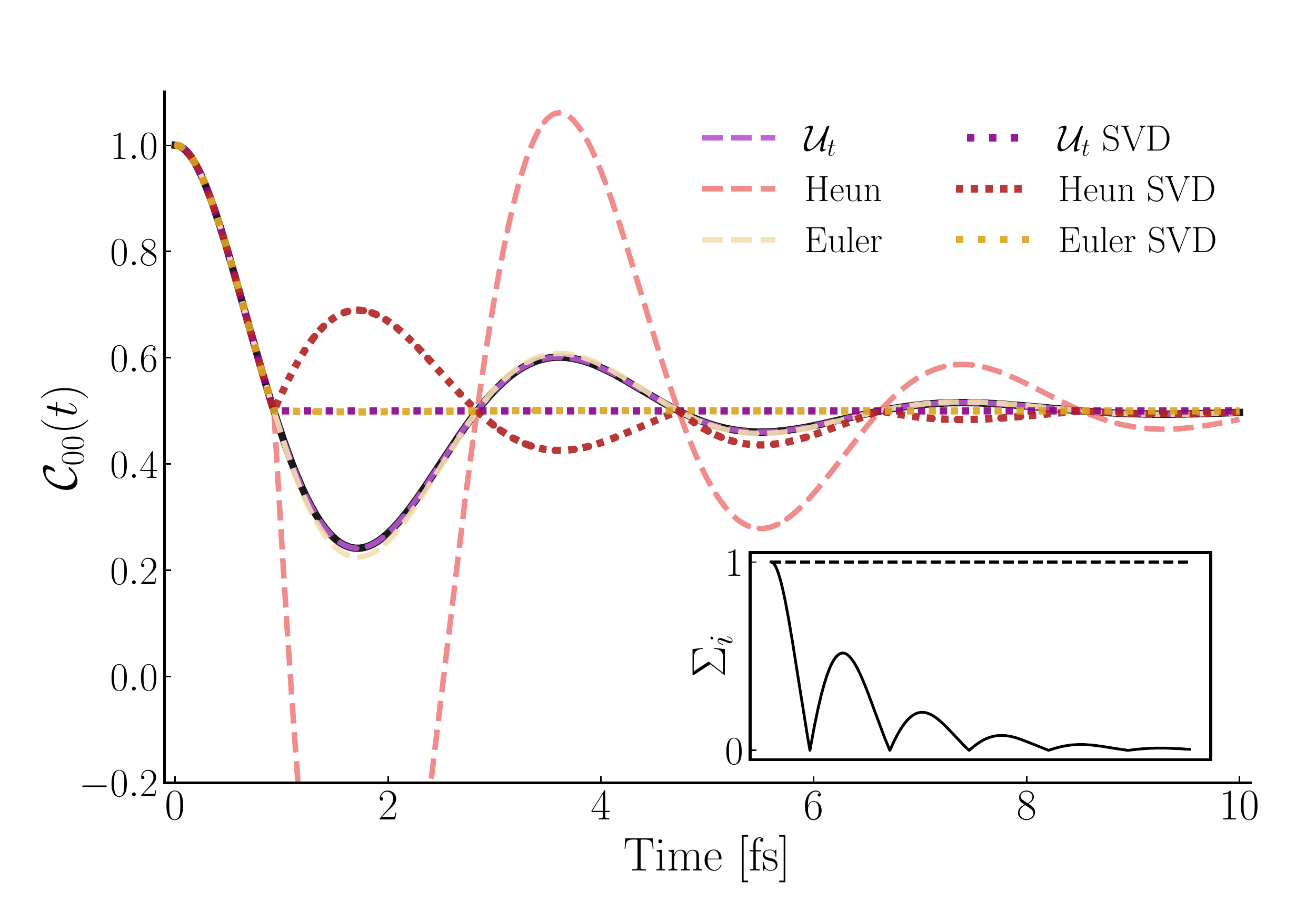}}\vspace{-2pt}
\end{center}
\vspace{-16pt}
\caption{\label{fig:avoid_pole} Populations-only projector on the unbiased model. Dashed red line is the same as in Fig.~\ref{fig:popsOnly_SB}, using Heun's method. Forming $\mathcal{R}(t)$ with the SVD of Eq.~\ref{eq:SVD} instead gives the red dotted line, which becomes out-of-phase at the first pole. In contrast, Euler's method gives smaller error (dashed yellow), but SVD quenches the dynamics at the pole (dotted yellow). The discrete-time TC-GME gives the same constant line when filtered with SVD (dotted purple), but gives perfect agreement without it (dashed purple). \textbf{Inset}: The singular values as function of time. The dashed line remains at unity, as pointed out in the main text.}
\end{figure}

For the Heun's method integrator used in the main text, this SVD introduces a sudden sign-change in the predicted dynamics, see Fig.~\ref{fig:avoid_pole}. This is because the sudden reduction in the magnitude of the $\mathcal{R}(t)$ object caused by removing the singular point is not compatible  with Heun's method, as it is a predictor-corrector integrator. One can find (by optimization) that $\mathcal{R}(t_\rmm{pole}/(-42)$ is the value that would actually lead to correct dynamics.

Perhaps, then, moving to a purely `time-local integrator' would remove this problem. In Fig.~\ref{fig:avoid_pole} we also show how Euler's method performs, both with and without the SVD filtering. Including the singular point, the error introduced by Euler's method for this pole is actually much smaller than the high order integrator. However, while this simple replacement of the integrator yields sufficiently accurate dynamics for this problem, it does not ensure that Euler's method provides a stable integrator of the GME in all applications. Interestingly, removing the singular point causes the dynamics to become suddenly quenched to $\mathcal{C}=\mathcal{C}_\rmm{eq}$. In this case the matrix reaches, rather than overshoots, the equilibrium value: further propagation of the equation of motion cannot move it out of the null space. 

It follows, then, that the discrete-time implementation of Sec.~\ref{sec:discrete-time-tcgme} will ignore this pathological spike because it never has to perform an integration. Not only does Fig.~\ref{fig:avoid_pole} show that this is the case, it also demonstrates that SVD on the $\mathcal{U}_t$ object (which has the same poles as $\mathcal{R}(t)$, by construction) also serves to trap the dynamics at the equilibrium values. This further demonstrates the superiority of the discrete-time writing of the TCL-GME.

\section{Rotated Time-local Generators}\label{app:Rrotated}
\begin{figure}[!b]
\vspace{-16pt}
\begin{center}
    \resizebox{.5\textwidth}{!}{\includegraphics{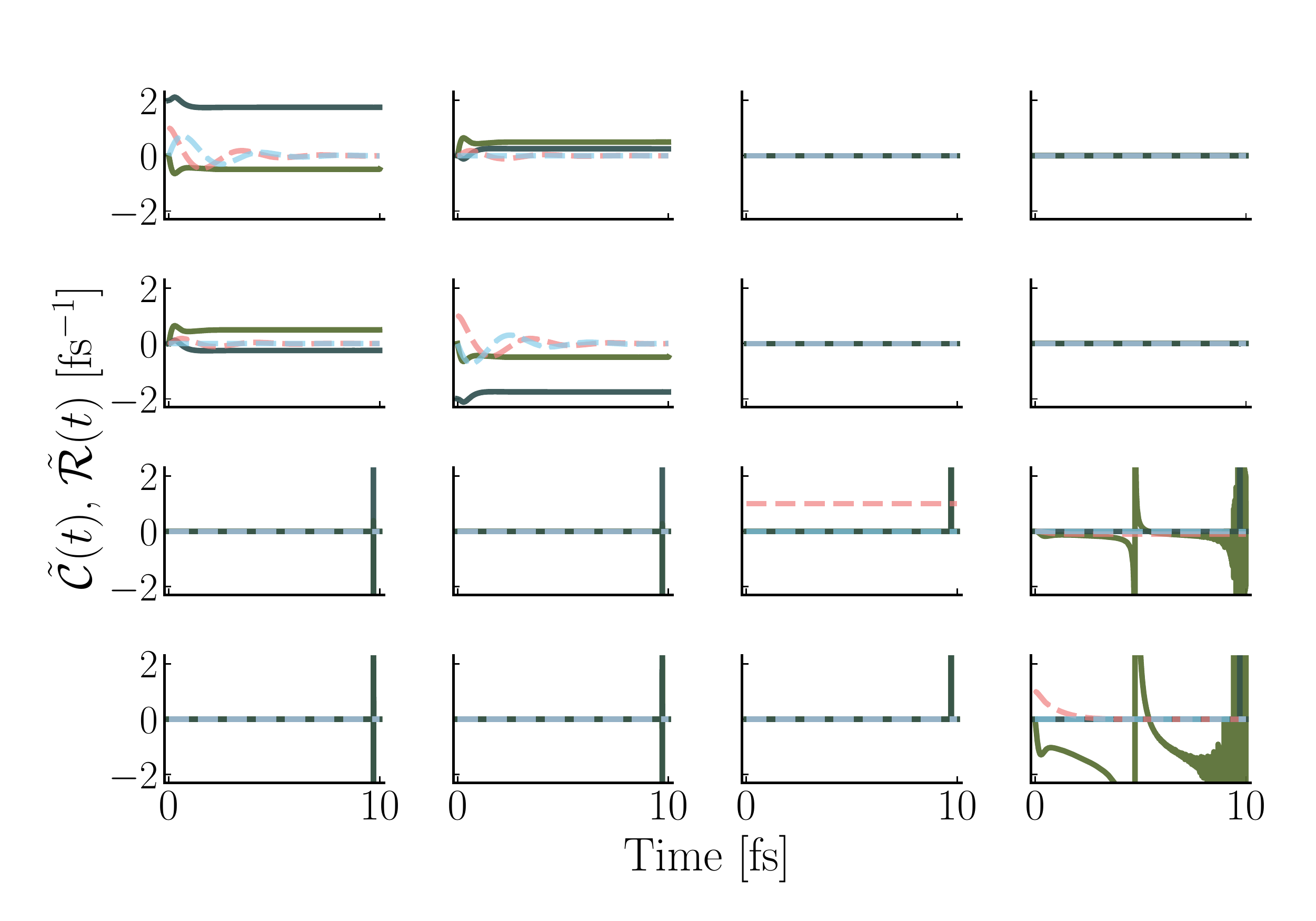}}
\end{center}
\vspace{-18pt}
\caption{\label{fig:rotated_unbiased} Full $\tilde{\mc{R}}(t)$ matrix for the unbiased spin boson problem, whose diagonal elements compose Fig.~\ref{fig:CR_rot_SB}. For the off-diagonal panel in the lower two-by-two block, $\tilde{\mc{C}}_{23}(t)$ begins at zero and smoothly plateaus to $-0.1$ within $2.5$~fs, while $\tilde{\mc{R}}_{23}(t)$ is $\simeq \tilde{\mc{R}}_{33}(t)/7$.}
\end{figure}

\begin{figure}[!b]
\vspace{-14pt}
\begin{center}
    \resizebox{.5\textwidth}{!}{\includegraphics{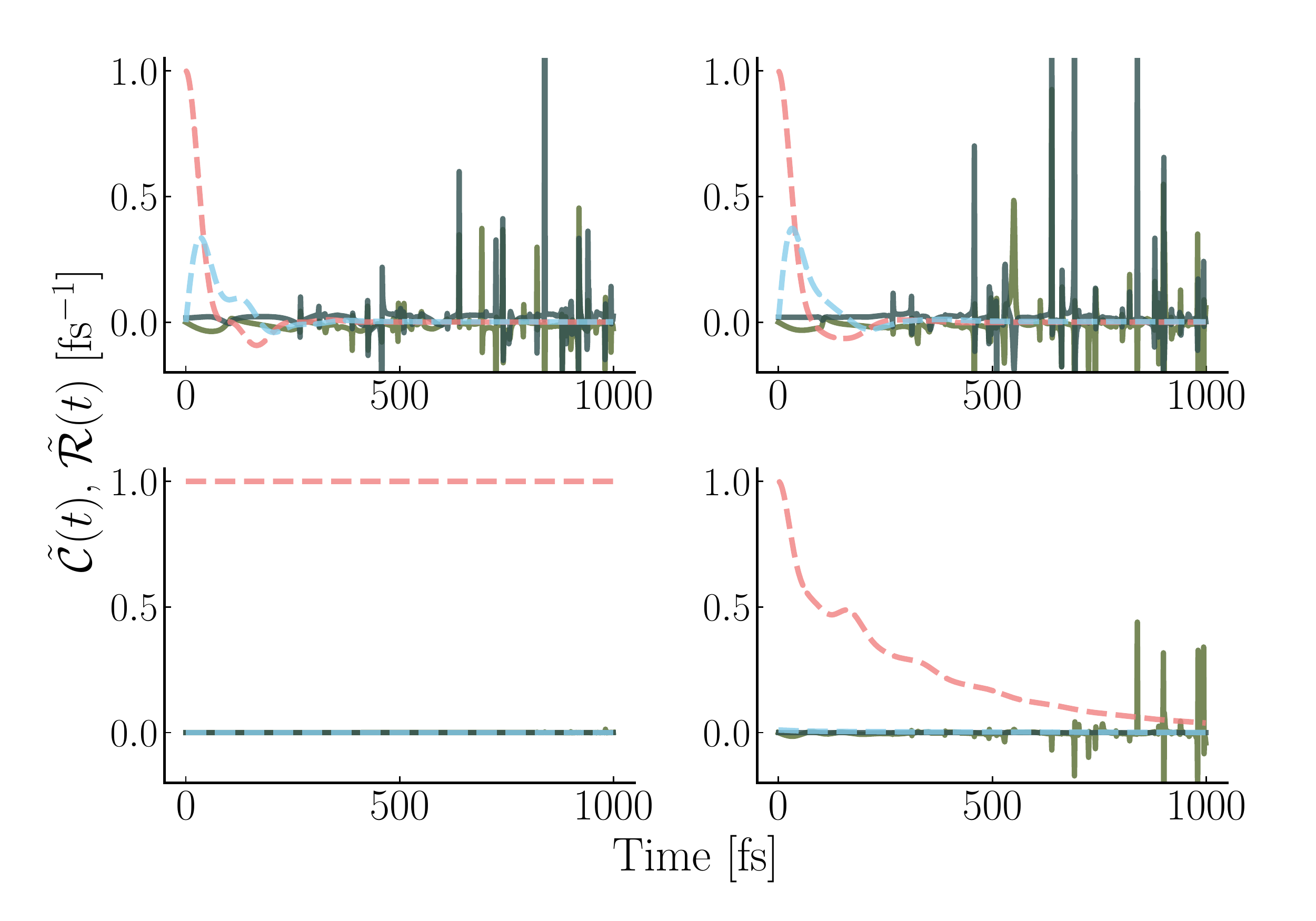}}
    \\
    \vspace{-0pt}
    \resizebox{.5\textwidth}{!}{\includegraphics{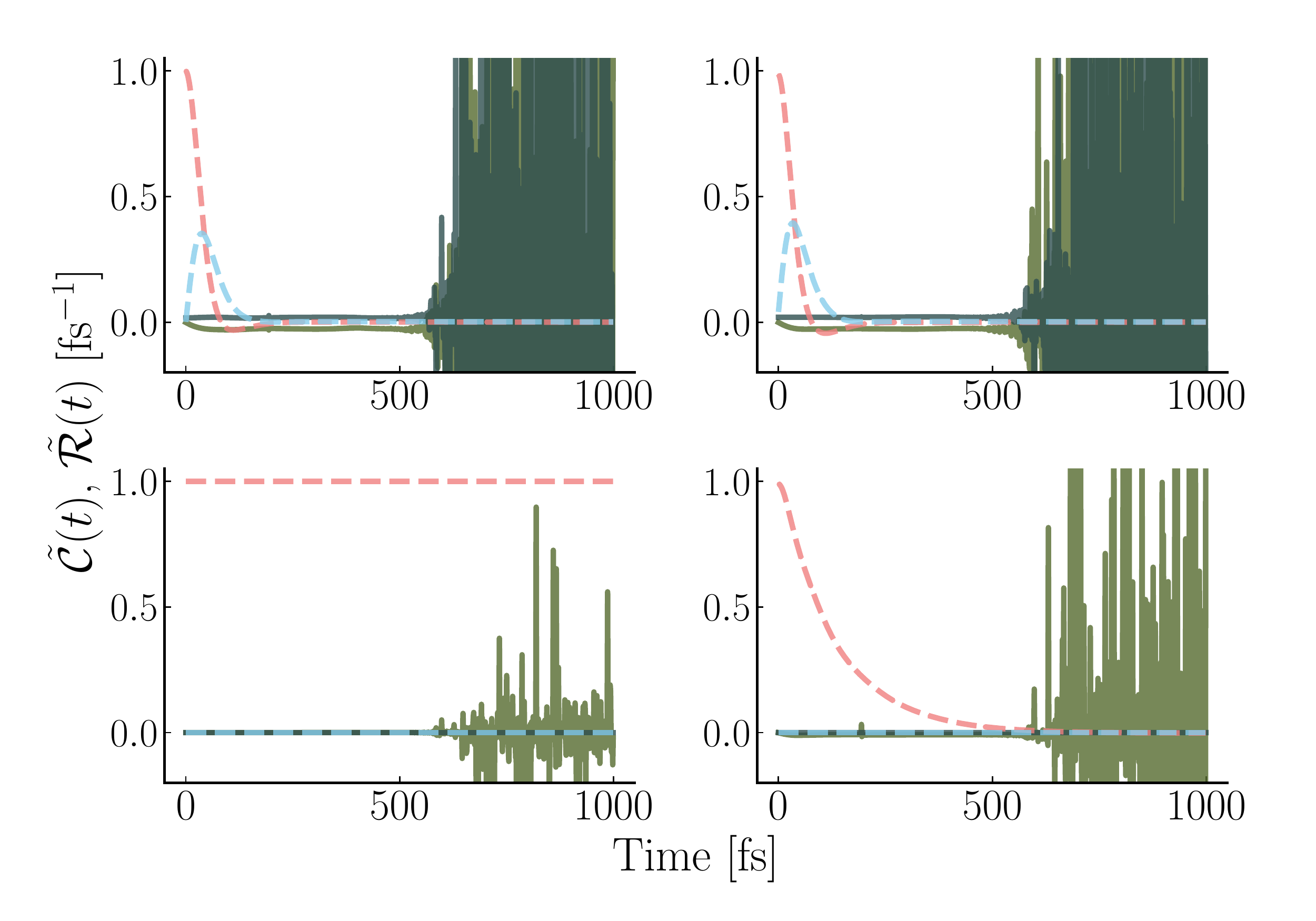}}
\end{center}
\vspace{-20pt}
\caption{\label{fig:rotated_baths} Representative elements of the rotated objects for the FMO model. $\tilde{\mc{R}}(t)$ in green and teal with $\tilde{\mc{C}}(t)$ in dashed red and blue; real and imaginary parts, respectively. These are the diagonal $ii$ elements for $i\in\{34, 37, 41, 44\}$ chosen to match the qualitative features of Fig.~\ref{fig:CR_rot_SB}. If $\tilde{\mc{R}}(t)$ is anywhere positive in the real part when truncated, the dynamics will diverge. \textbf{Top}: Slow bath with $\tau = 166$~fs where the onset of spikes occurs before any plateau. \textbf{Bottom}: Fast bath with $\tau = 50$~fs displaying plateau before the small spike around $200$~fs, which corresponds to the region of instability in Fig.~7.}
\end{figure}

We now turn to our analysis of the rotated dynamics and time-local generator, $\tilde{\mc{C}}(t)$ and $\tilde{\mc{R}}(t)$, in the eigenbasis of $\Omega$. Figures~\ref{fig:rotated_unbiased} and \ref{fig:rotated_baths} show (representative elements of) $\tilde{\mc{C}}(t)$ and $\tilde{\mc{R}}(t)$ for both the SB and FMO models discussed in the text.

Starting with the unbiased SB model, we show the full rotated  $\tilde{\mc{C}}(t)$ and $\tilde{\mc{R}}(t)$ in Fig.~\ref{fig:rotated_unbiased} corresponding to the parameter regime discussed in Figs.~\ref{fig:popsOnly_SB}, \ref{fig:CR_SB}, \ref{fig:CR_rot_SB} and \ref{fig:pathintegral_SB} in the main text. This figure illustrates the small off-diagonal contributions in the upper two-by-two block and the delocalized pole in $\tilde{\mc{R}}_{23}(t)$. In the main text, we neglected the contributions of \textit{all} off-diagonal entries. Finite off-diagonal elements are confined to the upper and lower diagonal blocks and can therefore only affect the diagonals in those same blocks. For the upper two-by-two there is no pole, and $\tilde{\mc{R}}(t)$ plateaus as expected, so the discussion is inconsequential to the analysis in the paper. For the lower two-by-two, the off-diagonal panel with the pole is qualitatively similar to the diagonal panel with the pole (but smaller, and with negative $\tilde{\mc{C}}(t)$). It therefore produces the same qualitative behavior of the lower-right panel when it multiplies into $\tilde{\mc{C}}_{33}(t)$; the conjugate element $\tilde{\mc{R}}_{32}(t)$ is null, and so $\tilde{\mc{R}}_{33}(t)$ in the bottom-right panel does not interact with any other dynamics and the analysis in the main text remains justified. In fact, since the panel $\tilde{\mc{C}}_{22}(t)=1~\forall~t$ is a statement of probability conservation, we suspect that this is an artifact and that more precise numerics may isolate the pole completely in the bottom-right panel.

We also show four representative elements for the two FMO models in Fig.~\ref{fig:rotated_baths}. In both cases, the majority of the diagonal matrix elements (of which there are 49) are qualitatively similar to those displayed in the first row of the figure, while the final~six are instead like the bottom-right panel; similarly each matrix has a single element $\tilde{\mc{C}}(t)=1~\forall~t$, both of which are displayed in the figure. Note that in these cases, the poles themselves are well-localized, and so $\tau_\rmm{R}$ ought to be determined by the plateau in $\tilde{\mc{R}}(t)$ alone.

\clearpage
\begin{figure}[!t]
\begin{center}
    \resizebox{.5\textwidth}{!}{\includegraphics{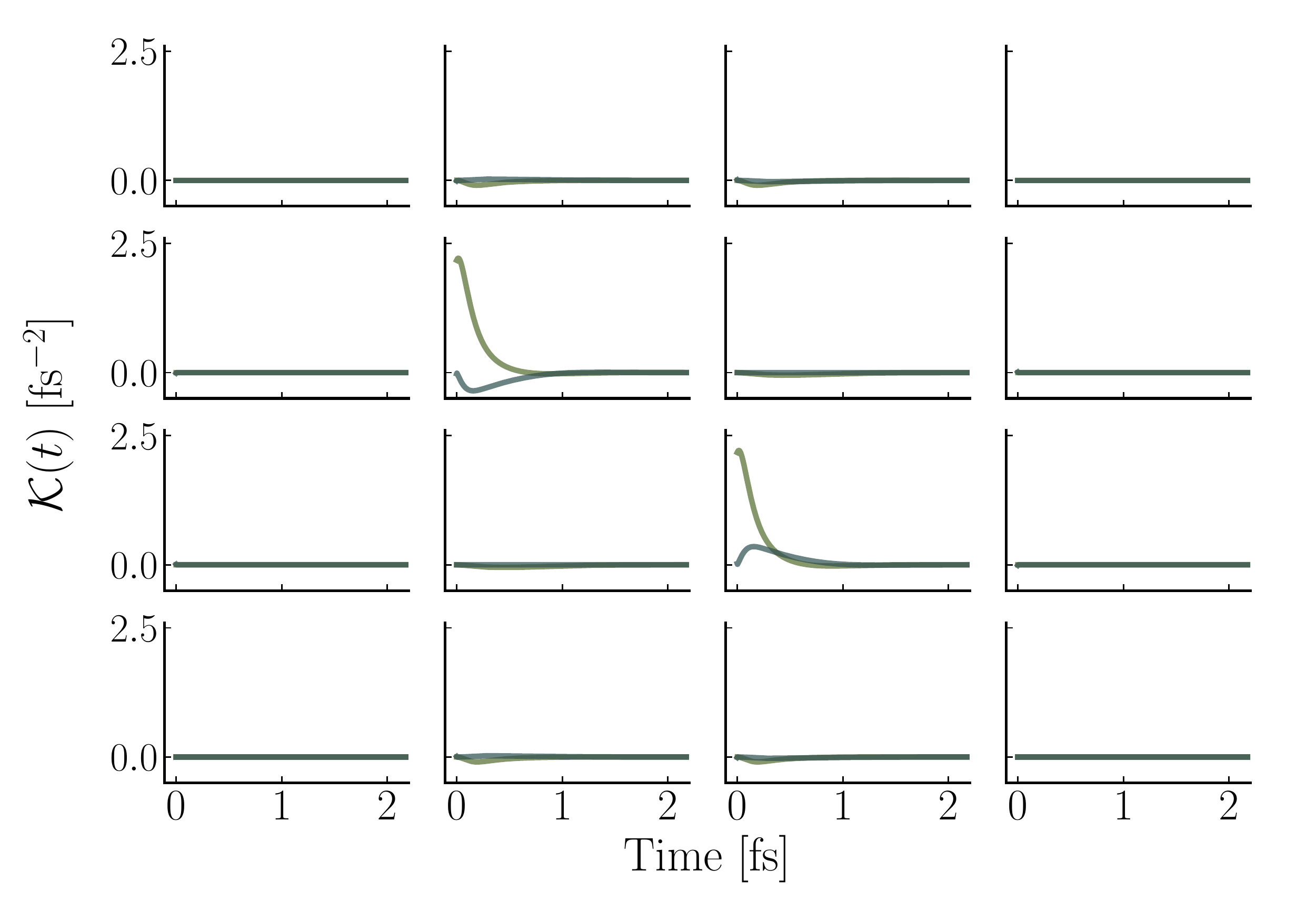}}\vspace{-2pt}
\end{center}
\vspace{-16pt}
\caption{\label{fig:memory_kernels_SB} Full memory kernel matrix for the biased spin-boson model used in this work. Note the `banded' structure where only the middle two columns are non-zero, to be compared with the time-local generator displayed in Fig.~\ref{fig:CR_SB}, top. Close inspection shows the kernel has decayed by $1.5$~fs.}
\end{figure}

\section{Time-nonlocal Kernels}\label{app:memorykernels}

For completeness, we provide representative elements from the time-nonlocal memory kernel, $\mc{K}(t)$, for the biased SB and FMO models discussed in the paper. Figure~\ref{fig:memory_kernels_SB} shows the memory kernel for the SB model with the Argyres-Kelly (full reduced electronic density matrix) projector. Here, the memory kernel decays by $\tau_{\mathcal{K}} \approx 1.5$~fs, in agreement with the chosen time-local generator cutoff in the manuscript. 

\begin{figure}[!h]
\begin{center}
    \resizebox{.5\textwidth}{!}{\includegraphics{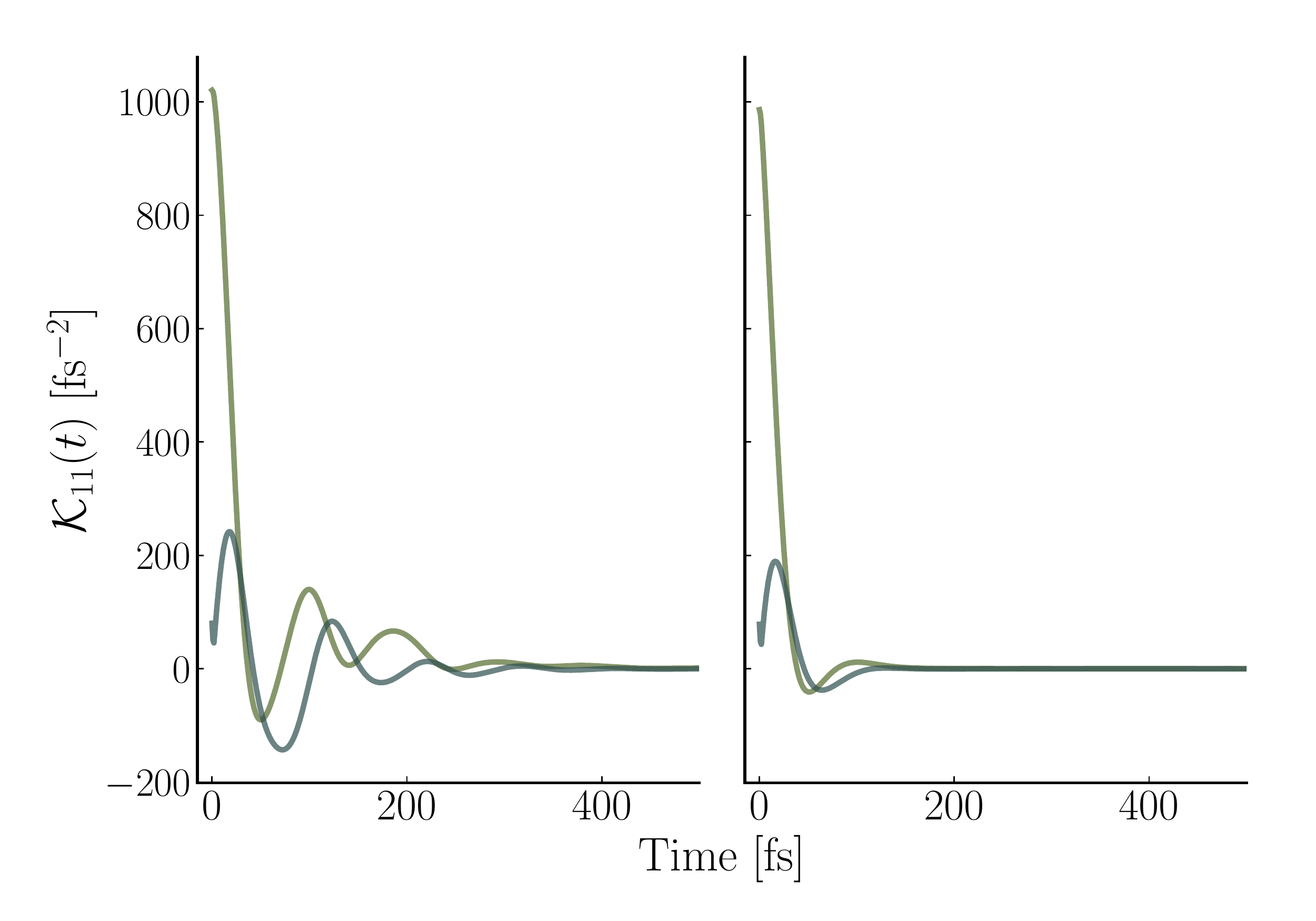}}\vspace{-2pt}
\end{center}
\vspace{-16pt}
\caption{\label{fig:memory_kernels_FMO} Representative memory kernel elements for the Frenkel exciton model parameterised to FMO. \textbf{Left}: Slow bath with $\tau = 166$~fs, showing longer-lived memory effects up to around $500$~fs. \textbf{Right}: Fast bath with $\tau = 50$~fs, showing that the system becomes Markovian before $200$~fs.}
\end{figure}

Figure~\ref{fig:memory_kernels_FMO} shows representative elements of the memory kernel for the Frenkel exciton model with slow and fast baths. Here, one can visually determine the kernel cutoff times to be $\tau_{\mathcal{K}}^{\rm slow} \approx 400$~fs and $\tau_{\mathcal{K}}^{\rm slow} \approx 150$~fs. The latter is in close agreement with $\tau_\mc{R}^{\rm fast}=160$~fs as we determined in Fig.~\ref{fig:CR_fast_FMO}, whereas the former is $170$~fs later than the last well-behaved guess for $\tau_\mc{R}^{\rm slow}$. This further supports the claim that the onset of spikes is faster than the time required to enter the Markovian regime for this slower bath.

\vfill


\bibliography{library, Postdoc-GQME}
\end{document}